\PassOptionsToPackage{unicode}{hyperref}
\PassOptionsToPackage{hyphens}{url}
\PassOptionsToPackage{dvipsnames,svgnames,x11names}{xcolor}
\documentclass[
]{article}

\usepackage{amsmath,amssymb}
\usepackage{iftex}
\ifPDFTeX
  \usepackage[T1]{fontenc}
  \usepackage[utf8]{inputenc}
  \usepackage{textcomp} 
\else 
  \usepackage{unicode-math}
  \defaultfontfeatures{Scale=MatchLowercase}
  \defaultfontfeatures[\rmfamily]{Ligatures=TeX,Scale=1}
\fi
\usepackage{lmodern}
\ifPDFTeX\else  
\fi
\IfFileExists{upquote.sty}{\usepackage{upquote}}{}
\IfFileExists{microtype.sty}{
  \usepackage[]{microtype}
  \UseMicrotypeSet[protrusion]{basicmath} 
}{}
\makeatletter
\@ifundefined{KOMAClassName}{
  \IfFileExists{parskip.sty}{%
    \usepackage{parskip}
  }{
    \setlength{\parindent}{0pt}
    \setlength{\parskip}{6pt plus 2pt minus 1pt}}
}{
  \KOMAoptions{parskip=half}}
\makeatother
\usepackage{xcolor}
\ifx\paragraph\undefined\else
  \let\oldparagraph\paragraph
  \renewcommand{\paragraph}[1]{\oldparagraph{#1}\mbox{}}
\fi
\ifx\subparagraph\undefined\else
  \let\oldsubparagraph\subparagraph
  \renewcommand{\subparagraph}[1]{\oldsubparagraph{#1}\mbox{}}
\fi

\usepackage{longtable,booktabs,array}
\usepackage{calc} 
\usepackage{etoolbox}
\makeatletter
\patchcmd\longtable{\par}{\if@noskipsec\mbox{}\fi\par}{}{}
\makeatother
\IfFileExists{footnotehyper.sty}{\usepackage{footnotehyper}}{\usepackage{footnote}}
\makesavenoteenv{longtable}
\usepackage{graphicx}
\makeatletter
\def\maxwidth{\ifdim\Gin@nat@width>\linewidth\linewidth\else\Gin@nat@width\fi}
\def\maxheight{\ifdim\Gin@nat@height>\textheight\textheight\else\Gin@nat@height\fi}
\makeatother
\setkeys{Gin}{width=\maxwidth,height=\maxheight,keepaspectratio}
\makeatletter
\def\fps@figure{htbp}
\makeatother
\newlength{\cslhangindent}
\newlength{\csllabelwidth}
\newlength{\cslentryspacingunit} 
\newenvironment{CSLReferences}[2] 
 {
  \setlength{\parindent}{0pt}
  \ifodd #1
  \let\oldpar\par
  \def\par{\hangindent=\cslhangindent\oldpar}
  \fi
  \setlength{\parskip}{#2\cslentryspacingunit}
 }%
 {}
\usepackage{calc}

\usepackage{booktabs}
\usepackage{longtable}
\usepackage{array}
\usepackage{multirow}
\usepackage{wrapfig}
\usepackage{float}
\usepackage{colortbl}
\usepackage{pdflscape}
\usepackage{tabu}
\usepackage{threeparttable}
\usepackage{threeparttablex}
\usepackage[normalem]{ulem}
\usepackage{makecell}
\usepackage{xcolor}
\usepackage{gensymb}
\usepackage{arxiv}
\usepackage{orcidlink}
\usepackage{amsmath}
\usepackage[T1]{fontenc}
\makeatletter
\@ifpackageloaded{caption}{}{\usepackage{caption}}
\AtBeginDocument{%
\ifdefined\contentsname
  \renewcommand*\contentsname{Table of contents}
\else
  \newcommand\contentsname{Table of contents}
\fi
\ifdefined\listfigurename
  \renewcommand*\listfigurename{List of Figures}
\else
  \newcommand\listfigurename{List of Figures}
\fi
\ifdefined\listtablename
  \renewcommand*\listtablename{List of Tables}
\else
  \newcommand\listtablename{List of Tables}
\fi
\ifdefined\figurename
  \renewcommand*\figurename{Figure}
\else
  \newcommand\figurename{Figure}
\fi
\ifdefined\tablename
  \renewcommand*\tablename{Table}
\else
  \newcommand\tablename{Table}
\fi
}
\@ifpackageloaded{float}{}{\usepackage{float}}
\floatstyle{ruled}
\@ifundefined{c@chapter}{\newfloat{codelisting}{h}{lop}}{\newfloat{codelisting}{h}{lop}[chapter]}
\floatname{codelisting}{Listing}

\makeatother
\makeatletter
\@ifpackageloaded{caption}{}{\usepackage{caption}}
\@ifpackageloaded{subcaption}{}{\usepackage{subcaption}}
\makeatother
\makeatletter
\@ifpackageloaded{tcolorbox}{}{\usepackage[skins,breakable]{tcolorbox}}
\makeatother
\makeatletter
\@ifundefined{shadecolor}{\definecolor{shadecolor}{rgb}{.97, .97, .97}}
\makeatother
\ifLuaTeX
  \usepackage{selnolig}  
\fi
\IfFileExists{bookmark.sty}{\usepackage{bookmark}}{\usepackage{hyperref}}
\IfFileExists{xurl.sty}{\usepackage{xurl}}{} 
\hypersetup{
  pdftitle={From pixels to parcels: flexible, practical small-area uncertainty estimation for spatial averages obtained from aboveground biomass maps},
  pdfauthor={Lucas K Johnson; Grant M Domke; Stephen V Stehman; Michael J Mahoney; Colin M Beier},
  pdfkeywords={model-based inference, bootstrap, Landsat, machine
learning, national forest inventory, Forest Inventory and Analysis},
  colorlinks=true,
  linkcolor={blue},
  filecolor={Maroon},
  citecolor={Blue},
  urlcolor={Blue},
  pdfcreator={LaTeX via pandoc}}

\renewcommand{\today}{2024-12-20}
\newcommand{\runninghead}{A Preprint }
\renewcommand{\runninghead}{From pixels to parcels }
\title{From pixels to parcels: flexible, practical small-area
uncertainty estimation for spatial averages obtained from aboveground
biomass maps}
\def\asep{\\\\\\ } 
\author{\textbf{Lucas K
Johnson}~\orcidlink{0000-0002-7953-0260}\\Department of Sustainable
Resources Management\\State University of New York College of
Environmental Science and Forestry\\Syracuse,
NY,\ 13210\\\href{mailto:johnsl27@oregonstate.edu}{johnsl27@oregonstate.edu}\asep\textbf{Grant
M Domke}~\orcidlink{0000-0003-0485-0355}\\Northern Research
Station\\USDA Forest Service\\St.~Paul,
MN,\ 55114\\\href{mailto:grant.m.domke@usda.gov}{grant.m.domke@usda.gov}\asep\textbf{Stephen
V Stehman}~\orcidlink{0000-0001-5234-2027}\\Department of Sustainable
Resources Management\\State University of New York College of
Environmental Science and Forestry\\Syracuse,
NY,\ 13210\\\href{mailto:svstehma@esf.edu}{svstehma@esf.edu}\asep\textbf{Michael
J Mahoney}~\orcidlink{0000-0003-2402-304X}\\Department of Sustainable
Resources Management\\State University of New York College of
Environmental Science and Forestry\\Syracuse,
NY,\ 13210\\\href{mailto:mjmahone@esf.edu}{mjmahone@esf.edu}\asep\textbf{Colin
M Beier}~\orcidlink{0000-0003-2692-7296}\\Department of Sustainable
Resources Management\\State University of New York College of
Environmental Science and Forestry\\Syracuse,
NY,\ 13210\\\href{mailto:cbeier@esf.edu}{cbeier@esf.edu}}
\date{2024-12-20}
\begin{document}
\maketitle
\begin{abstract}
Fine-resolution maps of forest carbon and aboveground biomass (AGB)
effectively represent spatial patterns and can be flexibly aggregated to
map subregions by computing spatial averages or totals of pixel-level
predictions. However, generalized model-based uncertainty estimation for
spatial aggregates requires computationally expensive processes like
iterative bootstrapping and computing pixel covariances. Uncertainty
estimation for map subregions is critical for enhancing practicality and
eventual adoption of model-based data products, as this capability would
empower users to produce estimates at scales most germane to management
of the landscape: individual forest stands and ownership parcels. In
this study we produced estimates of standard error (SE) associated with
spatial averages of fine-resolution AGB map predictions for a stratified
random sample of ownership parcels (0-5000 acres) in New York State
(NYS). This represents the first model-based uncertainty estimation
study to include all four types of uncertainty (reference data, sample
variability, residual variability, and auxiliary data), incorporate
spatial autocorrelation of model residuals, and use methods compatible
with algorithmic modeling (machine learning or nonparametric). We found
that uncertainty attributed to residual variance, largely resulting from
spatial correlation of residuals, dominated all other sources for the
majority of the parcels in the study (0-2500 acres). These results
suggest that improvements to model accuracy will yield the greatest
reductions to total uncertainty in regions like the northeastern and
midwestern United States where forests are divided into smaller spatial
units. Further, we demonstrated that log-log regression relating parcel
characteristics (area, perimeter, AGB density, forest cover) to
parcel-level SE can accurately estimate uncertainty for map subregions,
thus providing a convenient means to empower map users. These findings
support transparency in future regional-scale model-based forest carbon
accounting and monitoring efforts.
\end{abstract}
{\bfseries \emph Keywords}
\def\sep{\textbullet\ }
model-based inference \sep bootstrap \sep Landsat \sep machine
learning \sep national forest inventory \sep 
Forest Inventory and Analysis

\ifdefined\Shaded\renewenvironment{Shaded}{\begin{tcolorbox}[sharp corners, boxrule=0pt, enhanced, frame hidden, interior hidden, borderline west={3pt}{0pt}{shadecolor}, breakable]}{\end{tcolorbox}}\fi

\hypertarget{introduction}{%
\section{Introduction}\label{introduction}}

Over the past 30 years, greenhouse gas (GHG) accounting efforts have
proliferated from local to global scales and across private and public
sectors. These efforts began in 1992 with the United Nations requiring
all developed counties to report annual emissions ({``The United Nations
Framework Convention on Climate Change''} 1992), and since have evolved
into the Kyoto Protocol ({``Kyoto Protocol to the United Nations
Framework Convention on Climate Change''} 1997), and the Paris Climate
agreement ({``Paris Agreement to the United Nations Framework Convention
on Climate Change''} 2015) with accounting methodologies and practices
defined by the 2019 IPCC update on GHG inventory methods (Buendia et al.
2019). These standards require that GHG emissions and removals be
reported with confidence intervals and with uncertainties reduced as far
as possible (Penman et al. 2003; Eggleston et al. 2006). Beyond
satisfying these requirements, identifying and understanding sources of
uncertainty are critical for distinguishing between noise and real GHG
emissions and removals. Forests are among the most effective carbon
sinks with global net sequestration rates estimated at 1.1 (+/- 0.8) Pg
of carbon per year (Pan et al. 2011). However, the strength of this sink
is not spatially consistent and in some regions forest lands may be net
carbon sources due to deforestation and degradation (Pan et al. 2011).
Consequently, a significant body of research is dedicated to accounting
for forest carbon fluxes with improved accuracy, precision, and spatial
resolution.

Many large-scale forest carbon accounting approaches leverage the
extensive sampling designs of national forest inventories (NFI) to
estimate annual emissions and removals from the forest sector (Woodall
et al. 2015; McRoberts, Tomppo, and Næsset 2010). These design-based
estimators are often accompanied by relatively accessible estimators of
variance, which account for sampling variability but omit variability
related to tree- and plot-level predictions (reference data uncertainty)
since population units are assumed to have a single fixed value under
the design-based framework (McRoberts 2011; McRoberts et al. 2022).
Although design-based approaches offer fundamental insights and
essential data on forest carbon dynamics, they are often limited
spatially and temporally by sampling density and measurement frequency
respectively (McRoberts 2011), and thus cannot represent fine-scale
patterns and dynamics most relevant to planning and decision-making.
Model-based approaches, which combine field data (e.g.~NFI plots) with
wall-to-wall, remotely-sensed data can fill this need by producing
predictions for all map units (pixels) in a given area (Kennedy, Ohmann,
et al. 2018; Matasci et al. 2018; Huang et al. 2019; Hudak et al. 2020;
L. K. Johnson et al. 2022, 2023).

Although model-based approaches offer enhanced spatial flexibility
relative to their design-based counterparts, they rely on different
assumptions and require a more complex combination of terms for
estimating uncertainty. Under model-based inferential frameworks,
variability is derived from the assumption that each population unit has
an entire distribution of possible values, with each random realization
of a population unit value corresponding to a single distinct population
from an overarching set of populations known as a super-population
(McRoberts et al. 2022). Following model-based terminology,
super-population parameters are thus estimated, whereas population
parameters are predicted (McRoberts et al. 2022). Model-based estimates
of uncertainty may include the following components: uncertainty in
reference data (Table~\ref{tbl-uncertainty-types}, type a), uncertainty
due to training the model on a sample
(Table~\ref{tbl-uncertainty-types}, type b), uncertainty due to the
model not being able to represent all of the variation in the response
variable (Table~\ref{tbl-uncertainty-types}, type c), and auxiliary data
uncertainty (Table~\ref{tbl-uncertainty-types}, type d) (Wadoux and
Heuvelink 2023; CEOS 2021). Different terms have been used to refer to
type b, including model parameter/prediction covariance, model
parameter/prediction uncertainty (McRoberts et al. 2018), or sampling
variability (CEOS 2021; McRoberts et al. 2022) since model parameter
estimates and subsequent predictions depend on the training data sample.
Type c is usually called residual variability and is sometimes broken
down into components quantifying residual variance and spatial
covariance of residuals (McRoberts et al. 2018, 2022; CEOS 2021). The
proportion of model-based studies that produce uncertainty estimates is
small (McRoberts 2011), likely because the technical complexity of
generating uncertainty estimates is substantially greater than that of
making predictions.

From the model-based studies that do include estimates of uncertainty,
desirable features for the uncertainty estimation process include:

\begin{enumerate}
\def\labelenumi{\arabic{enumi}.}
\item
  Flexibility to incorporate parametric as well as algorithmic models
  (i.e. machine learning and nonparametric models).
\item
  Incorporating all four types of uncertainty
  (Table~\ref{tbl-uncertainty-types}).
\item
  Methods for producing uncertainty estimates for spatial averages or
  totals that explicitly account for spatial autocorrelation of
  residuals.
\item
  Consideration of computational efficiency and delivery to map users.
\end{enumerate}

Applications of rigorous model-based uncertainty estimation have been
advanced using parametric models with accompanying analytic estimators
of variance (Saarela et al. 2020, 2018, 2016; Chen et al. 2016; Chen,
Laurin, and Valentini 2015; McRoberts et al. 2018). However, these
analytic estimators are fairly complex, and can be inaccessible for
practitioners lacking deep backgrounds in statistics or mathematics.
Further, requiring parametric models may be limiting, and thus in
violation of desirable criterion 1 (defined above), as algorithmic
approaches may yield improved predictive accuracy (Efron 2020; Breiman
2001b). At the cost of computational efficiency, iterative approaches
like bootstrapping or Monte Carlo simulations, however, are flexible to
model form, can easily incorporate uncertainty from several sources, and
are conceptually simple relative to analytic estimators (McRoberts et
al. 2022; CEOS 2021; Esteban et al. 2020).

Reference data uncertainty (Table~\ref{tbl-uncertainty-types}, type a)
is often deemed negligible or is omitted due to a scarcity of
representative data describing measurement error and allometric
uncertainty (Breidenbach et al. 2014), thus violating desirable
criterion 2 (defined above). For forest aboveground biomass (AGB) or
carbon studies in particular, reference data may be highly uncertain,
given that tree-level AGB values in NFIs are typically predictions
themselves. Instead, tree-level biomass and carbon are modeled as a
function of measured variables like diameter and height using allometric
models developed for particular regions and taxa (Jenkins et al. 2003;
Woodall et al. 2011; Chave et al. 2014). The measured variables
themselves are subject to error and imprecision, further contributing to
the overall uncertainty of the tree-level predictions (Berger et al.
2014; Yanai et al. 2023). It is largely agreed that tree-level
uncertainty contributes a small fraction of the total uncertainty when
aggregated over many individuals for design-based estimates at coarse
scales (Breidenbach et al. 2014; Ståhl et al. 2014; McRoberts and
Westfall 2014; Yanai et al. 2023), yet whether or not these small
fractions can be considered negligible across scales and contexts is
unresolved (Chen, Laurin, and Valentini 2015; McRoberts et al. 2016;
Saarela et al. 2020). Considering the range of reference data
uncertainty contributions reported in the model-based estimation
literature (5\% of pixel-level uncertainty Chen, Laurin, and Valentini
(2015); 75\% of 5005 km\textsuperscript{2} area uncertainty Saarela et
al. (2020)), it remains prudent to include this component to the extent
that it is feasible.

Model-based estimates of pixel- or point-level uncertainty
(i.e.~uncertainty maps), though helpful to identify the spatial
distribution of uncertainties within an area, should be accompanied with
methods to aggregate uncertainties to produce estimates at scales
relevant to management and decision making like individual forest stands
and ownership parcels (desirable criterion 3, defined above). Spatially
aggregating predictions can be achieved by averaging or summing all
predictions within a given area. However, spatially aggregating
uncertainties requires data beyond that which is contained in an
uncertainty map (Wadoux and Heuvelink 2023; McRoberts et al. 2022).
Specifically, pairwise prediction or residual covariances and associated
spatial correlations are required but do not fit into an uncertainty map
(N\textsuperscript{2} pairs but N map pixels) and may present obstacles.
First, computing the necessary prediction or residual covariances for a
given area is often computationally expensive; second, obtaining the
spatially dense reference data (i.e.~short distances between plots)
needed to incorporate spatial autocorrelation of residuals is often
prohibitive (Breidenbach, McRoberts, and Astrup 2016; McRoberts et al.
2022; Wadoux and Heuvelink 2023). If estimates are only to be produced
for large units of aggregation (e.g.~polygons with dimensions well
beyond the range of spatial autocorrelation) then the contribution from
spatially autocorrelated residuals is negligible, however, for scales
germane to forest management (i.e. stands and parcels) this component
may be highly relevant (Breidenbach, McRoberts, and Astrup 2016;
McRoberts et al. 2018, 2022).

McRoberts et al. (2022) considered the feasibility of efficiently
producing uncertainty estimates for map subregions, both internally as a
map producer but also for map users in satisfaction of desirable
criterion 4 (defined above). Their approach was predicated on methods
for reducing the size of the metadata needed to estimate uncertainty for
map subregions to transfer this metadata more easily to users. However,
McRoberts et al. (2022) applied their proposed approach only to
relatively large areas (2500 km\textsuperscript{2} and 12,500
km\textsuperscript{2}) for which the effects of spatial autocorrelation
of residuals could be ignored. Further, this approach requires a
sophisticated series of operations including the reclassification of
pixel-predictions into bins, a lookup of bin covariances in a provided
matrix for each pixel pair, and summation of covariances for all pixel
pairs in the region of interest.

In this study we built on the methods developed in McRoberts et al.
(2022) to produce model-based estimates of uncertainty (standard error;
SE) associated with spatial averages of AGB predictions for small
ownership parcels (spatial units of analysis sized \textless1-5000
acres) in New York State (NYS). We replicated the machine learning
modeling framework developed in L. K. Johnson et al. (2023) to produce
pixel-level AGB predictions. We had three objectives: 1) to estimate a
model-based SE that included all four types of uncertainty
(Table~\ref{tbl-uncertainty-types}, types a-d), incorporated spatial
autocorrelation of model residuals, and used methods that are compatible
with algorithmic modeling; 2) estimate the relative contributions of
uncertainty components across scales to help direct further efforts
toward reducing total uncertainties in model-based estimates; and 3)
derive a relationship between ownership parcel characteristics (area,
perimeter, AGB density, \% forest cover) and SE in the form of a
multiple regression model to provide a practical, broadly applicable
method for uncertainty estimation for map subregions that both map users
and producers can implement, satisfying desirable criterion 4.

\hypertarget{tbl-uncertainty-types}{}
\begin{table}
\caption{\label{tbl-uncertainty-types}Categorization of four types of model-based uncertainty considered in
this study, and how they contribute (variance component) to the
estimates of total variance (\(\operatorname{Var_{total}}\)). }\tabularnewline

\centering
\begin{tabular}[t]{ll>{\raggedright\arraybackslash}p{15em}l}
\toprule
Type & Name & Source of Uncertainty & Variance Component\\
\midrule
a & Reference data uncertainty ($\operatorname{Var_{ref}}$) & Measurement, classification, allometric model, and plot location errors & $\operatorname{Var_{boot}}$\\
\addlinespace
b & Sampling variability ($\operatorname{Var_{sam}}$) & Training the model on a sample & $\operatorname{Var_{boot}}$\\
\addlinespace
c & Residual variability & The model is not able to represent all variation in the response variable & $\operatorname{Var_{res}}$\\
\addlinespace
d & Auxiliary data uncertainty ($\operatorname{Var_{LC}}$) & Variability related to sensor imprecision and inaccuracy or model uncertainty if the data are model generated & $\operatorname{Var_{boot}}$\\
\bottomrule
\end{tabular}
\end{table}

\hypertarget{data-and-methods}{%
\section{Data and Methods}\label{data-and-methods}}

\hypertarget{overview}{%
\subsection{Overview}\label{overview}}

\begin{figure}

{\centering \includegraphics{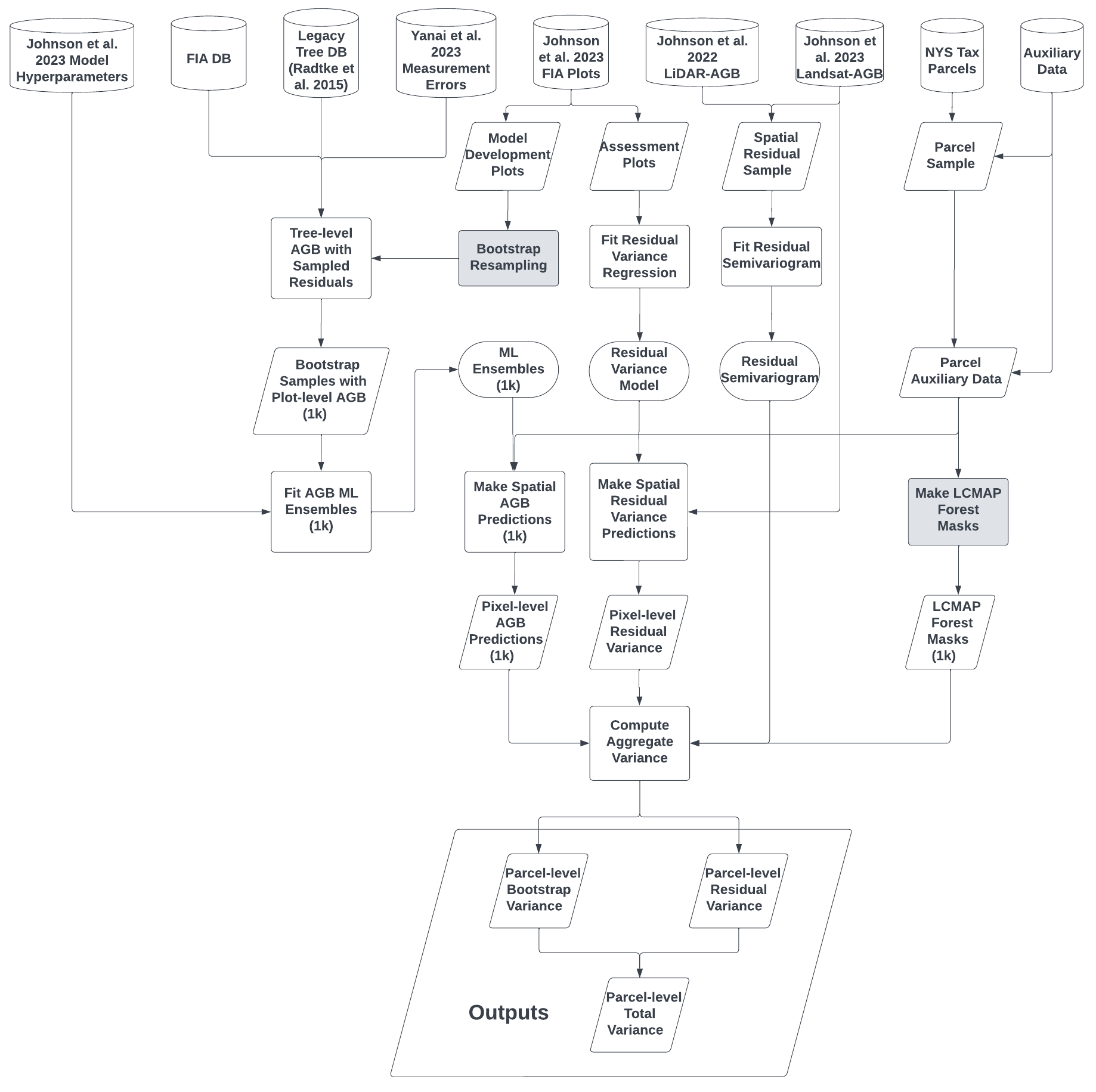}

}

\caption{\label{fig-flow}A flowchart showing the key elements of the
methodology to estimate uncertainty for spatial means of aboveground
biomass. Cylinders represent data sources, parallelograms represent data
products and results, rectangles represent processing steps, and ovals
represent models. Gray shaded nodes represent components that were
omitted for separate runs of the workflow to further deconstruct
\(\operatorname{Var_{boot}}\) by subtraction into
\(\operatorname{Var_{sam}}\), \(\operatorname{Var_{LC}}\), and
\(\operatorname{Var_{ref}}\).}

\end{figure}

We largely followed the steps described in McRoberts et al. (2022) to
produce estimates of total variance and standard error (SE) associated
with spatial averages of aboveground biomass (AGB) predictions for a
stratified random sample of small ownership parcels (\textless1-5000
acres) in New York State (NYS; Figure~\ref{fig-flow}). Pixel-level AGB
predictions were made with a machine learning ensemble model developed
in L. K. Johnson et al. (2023) (``direct'' approach) which relied on
Landsat imagery and was trained on plot-level AGB predictions from the
United States Forest Inventory and Analysis (FIA) program (Bechtold and
Patterson 2005; Gray et al. 2012). The total variance for a given parcel
was estimated as the sum of two components, and the SE was computed as
the square root of total variance as follows:

\begin{equation}\protect\hypertarget{eq-vartot}{}{
\operatorname{Var_{total} = Var_{boot} + Var_{res}}
}\label{eq-vartot}\end{equation}

\begin{equation}\protect\hypertarget{eq-se}{}{
\operatorname{SE} = \sqrt{Var_{total}}
}\label{eq-se}\end{equation}

Implicit in these formulas is the assumption that components
\(\operatorname{Var_{boot}}\) are \(\operatorname{Var_{res}}\) are
independent, but interrogating this assumption was beyond the scope of
this study.

After computing \(\operatorname{Var_{total}}\) and SE for each parcel in
our sample, we further deconstructed the \(\operatorname{Var_{boot}}\)
component and evaluated the relative contributions of each sub-component
to the total. Further, we fit a regression model to estimate SE for a
given parcel as a function of the following parcel characteristics:
area, perimeter, AGB, and forest cover. This regression model was
developed to expedite the process of uncertainty estimation for any
specified subregion of interest. All data compilation, modeling,
analysis, and summary were conducted using the R programming language (R
Core Team 2023) with the targets pipeline tool (Landau 2021).

\hypertarget{sec-boot-var}{%
\subsection{Bootstrap variance}\label{sec-boot-var}}

This component includes uncertainty in reference data
(Table~\ref{tbl-uncertainty-types}, type a), uncertainty due to training
a model on a sample (Table~\ref{tbl-uncertainty-types}, type b), and
uncertainty related to the landcover mask used in our areal estimates
(Table~\ref{tbl-uncertainty-types}, type d). The same iterative
bootstrapping approach was used to include each of these three types of
uncertainty. Specifically, in each iteration \emph{i} of 1000
iterations, a new training data set was developed as follows (additional
details for some steps are provided in the subsections noted in
parentheses):

\begin{enumerate}
\def\labelenumi{\arabic{enumi}.}
\item
  A sample was selected with replacement from the original sample of FIA
  field plots.
\item
  New tree-level AGB predictions were produced for all trees in the new
  sample of plots. Residuals were drawn from both empirical and
  estimated distributions to account for measurement error and
  allometric uncertainty/error in tree-level biomass predictions
  (Section~\ref{sec-tree-err}; Section~\ref{sec-allom-err}).
\item
  These new tree-level AGB predictions were aggregated within each plot
  to produce plot-level AGB in units of megagrams per hectare
  \(\operatorname{Mg\ ha^{-1}}\).
\item
  A residual was randomly selected from an estimated distribution of
  plot-location errors yielding a new plot geometry from which auxiliary
  data were extracted (Section~\ref{sec-loc-err};
  Section~\ref{sec-aux-data}).
\end{enumerate}

Next, machine learning (ML) ensembles were fit to each training data set
\emph{i}, and then used to make a set of predictions for the map units
(pixels) within each parcel, resulting in 1000 predictions per pixel
(Section~\ref{sec-agbmodelmap}). The bootstrap variance for a given
parcel was then computed as the variance over 1000 replications of the
parcel-level AGB density, where AGB density was computed as the average
of all pixel-predictions within the parcel. Specifically, the bootstrap
variance was computed with the following equation:

\begin{equation}\protect\hypertarget{eq-varboot}{}{
\operatorname{Var_{boot}} = \frac{\sum_{i=1}^n{(\hat{u_i} - \hat{u})^2}}{n - 1}
}\label{eq-varboot}\end{equation}

where \(n\) is the number of bootstrap iterations, \(\hat{u}\) is the
average parcel-level AGB density over \(n\) iterations, and
\(\hat{u_i}\) is the parcel-level AGB density estimated for a given
bootstrap iteration \emph{i} and computed as:

\begin{equation}\protect\hypertarget{eq-bootmean}{}{
\hat{u_i} = \frac{\sum_{i=1}^N{\hat{y_i}}}{N}
}\label{eq-bootmean}\end{equation}

where \(N\) is the number of pixels in a given parcel and \(\hat{y_i}\)
is a pixel-level AGB prediction made by the ML ensemble. Derivations for
and examples of these bootstrap procedures and equations can be found in
Freedman (1981), Efron and Tibshirani (1994), CEOS (2021), and McRoberts
et al. (2022).

To better understand the relative contributions of various sources of
uncertainty, we further deconstructed \(\operatorname{Var_{boot}}\) into
reference data variance (\(\operatorname{Var_{ref}}\);
Table~\ref{tbl-uncertainty-types}, type a), sample variance
(\(\operatorname{Var_{sam}}\); Table~\ref{tbl-uncertainty-types}, type
b), and landcover mask variance (\(\operatorname{Var_{LC}}\);
Table~\ref{tbl-uncertainty-types}, type d) such that:

\begin{equation}\protect\hypertarget{eq-bootdecon}{}{
\operatorname{Var_{boot}} = \operatorname{Var_{ref}} + \operatorname{Var_{sam}} + \operatorname{Var_{LC}}
}\label{eq-bootdecon}\end{equation}

Each respective sub-component was computed by subtraction. To compute
\(\operatorname{Var_{sam}}\), a 1000 iteration simulation was run with a
constant training data set across each iteration. The resulting estimate
of \(\operatorname{Var_{boot}}\), identified here as
\(\operatorname{Var_{boot-no-resample}}\), was then subtracted from the
original estimate of \(\operatorname{Var_{boot}}\) to produce
\(\operatorname{Var_{sam}}\) as follows:

\begin{equation}\protect\hypertarget{eq-varsam}{}{
\operatorname{Var_{sam}} = \operatorname{Var_{boot}} - \operatorname{Var_{boot-no-resample}}
}\label{eq-varsam}\end{equation}

Similarly, to compute \(\operatorname{Var_{LC}}\), a 1000 iteration
simulation was run with a constant landcover mask
(Section~\ref{sec-agbmodelmap}) across each iteration. The resulting
estimate of \(\operatorname{Var_{boot}}\), identified here as
\(\operatorname{Var_{boot-static-LC}}\), was then subtracted from the
original estimate of \(\operatorname{Var_{boot}}\) to produce
\(\operatorname{Var_{LC}}\) as follows:

\begin{equation}\protect\hypertarget{eq-varlc}{}{
\operatorname{Var_{LC}} = \operatorname{Var_{boot}} - \operatorname{Var_{boot-static-LC}}
}\label{eq-varlc}\end{equation}

In cases where the sum of \(\operatorname{Var_{sam}}\) and
\(\operatorname{Var_{LC}}\) was greater than
\(\operatorname{Var_{boot}}\), the difference between
\(\operatorname{Var_{boot}}\) and the sum of
\(\operatorname{Var_{sam}}\) and \(\operatorname{Var_{LC}}\) was
proportionally subtracted from each variance component
(\(\operatorname{Var_{sam}}\), \(\operatorname{Var_{LC}}\)). Finally,
estimates of \(\operatorname{Var_{sam}}\) and
\(\operatorname{Var_{LC}}\) were subtracted from original estimates of
\(\operatorname{Var_{boot}}\) to produce \(\operatorname{Var_{ref}}\) as
follows:

\begin{equation}\protect\hypertarget{eq-varref}{}{
\operatorname{Var_{ref}} = \operatorname{Var_{boot}} - \operatorname{Var_{sam}} - \operatorname{Var_{LC}}
}\label{eq-varref}\end{equation}

\hypertarget{sec-res-var}{%
\subsection{Residual variance}\label{sec-res-var}}

This component contributes to uncertainty due to the model failing to
represent all of the variation in the response variable
(Table~\ref{tbl-uncertainty-types}, type c). Because model residuals had
non-constant variance, we fit a model to estimate residual variance as a
function of our AGB predictions (Section~\ref{sec-resvarmod}). We used
this model to make pixel-level residual variance estimates that aligned
with individual AGB predictions. We then aggregated pixel-level residual
variance to each parcel as follows:

\begin{equation}\protect\hypertarget{eq-varres}{}{
\operatorname{Var_{res}} = \frac{\sum_{i=1}^N{\hat{\sigma_i}^{2}}}{N^2} + \frac{\sum_{i=1,i\neq j}^N{\sum_{j=1}^N}{\hat{\sigma_i} \hat{\sigma_j} \hat{\rho_{ij}}}}{N^2}
}\label{eq-varres}\end{equation}

where \(\hat{\sigma_i}^{2}\) is the estimated residual variance for a
given pixel, \(N\) is the number of pixels in a given parcel, and
\(\hat{\rho_{ij}}\) is the estimate of spatial correlation between
pixel-level residuals defined as

\begin{equation}\protect\hypertarget{eq-varrho}{}{
\hat{\rho_{ij}} = \frac{sill - \gamma(h)}{sill}
}\label{eq-varrho}\end{equation}

where \(\gamma\) is a semivariogram function, \(h\) is the distance
between pixels \(i\) and \(j\), and \(sill\) is the semivariance ceiling
estimated from the semivariogram. Derivations for
Equation~\ref{eq-varres} can be found in McRoberts et al. (2022), and
derivations for Equation~\ref{eq-varrho} can be found in Wadoux and
Heuvelink (2023) and Webster and Oliver (2007).

The semivariogram function used in Equation~\ref{eq-varrho} (further
described in Section~\ref{sec-spatcormeth}) was fit to a simple random
sample of spatial residuals (pixels) computed as follows:

\begin{equation}\protect\hypertarget{eq-residuals}{}{
\hat{e_i} = \hat{y}_{i-Landsat} - \hat{y}_{i-LiDAR}
}\label{eq-residuals}\end{equation}

where \(\hat{y}_{i-Landsat}\) is an AGB prediction from the
Landsat-based model evaluated in this study
(Section~\ref{sec-agbmodelmap}; L. K. Johnson et al. (2023)), and
\(\hat{y}_{i-LiDAR}\) is an AGB prediction from a LiDAR-based model
developed in L. K. Johnson et al. (2022).

\hypertarget{sec-studyarea}{%
\subsection{Study area and parcel sample}\label{sec-studyarea}}

NYS covers 141,297 km\textsuperscript{2} in the northeastern US and was
approximately 59\% forested as of 2019 (USFS 2020). The forests are
dominated by northern hardwoods-hemlock types but include Appalachian
oak and beech-maple-basswood forests in the western and southern regions
of the state respectively (Dyer 2006). The majority of NYS forests are
privately owned (\textasciitilde73\%; USDA Forest Service (2020)) and
contained within parcels smaller than 40 ha (\textasciitilde65\%; L'Roe
and Allred (2013)). This already fragmented forest landscape is
increasingly undergoing subdivision (L'Roe and Allred 2013).

\begin{figure}

{\centering \includegraphics{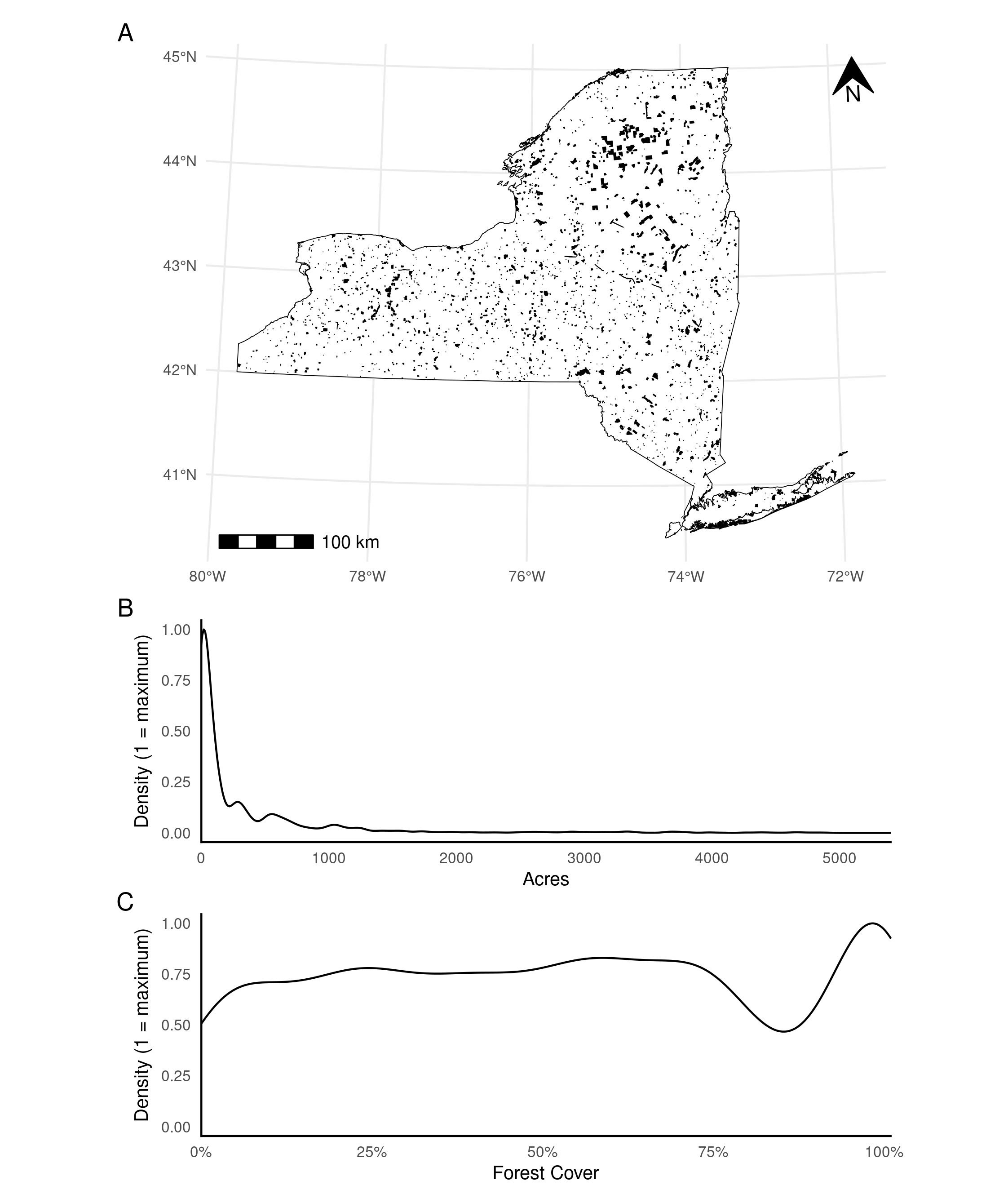}

}

\caption{\label{fig-parcelsummary}Parcel sample summary. A) Spatial
distribution of sampled parcels within New York State. B) Smoothed
frequency distribution of parcel sizes (acres) sampled. C) Smoothed
frequency distribution of forest cover (\%) sampled. Distributions for B
and C have been rescaled separately, such that the most common
occurrences of each variable are assigned a value of 1.}

\end{figure}

To limit computational scope we selected a stratified random sample of
ownership parcels within NYS. The statewide parcel database, compiled
for property taxation, was obtained under agreement with the NYS ITS
Geospatial Services. We stratified the statewide parcels into ten groups
based on size, and five groups based on percent forest cover as
determined by the United States Geological Survey's Land Change
Monitoring, Assessment, and Projection (LCMAP) version 1.2 primary
classification products (Zhu and Woodcock 2014; Brown et al. 2020). Size
strata were selected to ensure a sample that represented a wide range of
parcels, with an emphasis on smaller parcel sizes that dominate the NYS
ownership landscape and are underrepresented in uncertainty estimation
studies. Size strata were constructed as intervals (left-exclusive,
right-inclusive) of parcel acreage as follows: (0, 5{]}, (5, 10{]}, (10,
20{]}, (20, 50{]}, (50, 100{]}, (100, 250{]}, (250, 500{]}, (500,
1000{]}, (1000, 2500{]}, (2500, 5000{]}. Forest cover strata were
selected such that a range of forest conditions were represented within
each size stratum. Forest cover strata were constructed as five equal
intervals from 0 (exclusive) to 100 (inclusive) percent. Forest cover
was defined as the combination of the tree cover, wetlands, and
grass/shrub LCMAP classes (L. K. Johnson et al. 2023). 50 parcels were
randomly sampled from each size and cover group. For groups without 50
parcels, all available parcels were selected. Overall, 2224 parcels were
sampled across the state (Figure~\ref{fig-parcelsummary}).

\hypertarget{sec-field-data}{%
\subsection{Field data}\label{sec-field-data}}

The NYS FIA inventory data compiled in L. K. Johnson et al. (2023) were
partitioned into model development (80\%) and map assessment (20\%)
sets. The model development set was further partitioned into an 80\%
training set, and a 20\% validation used for iterative assessment during
model development. We used the model development partition to compute
\(\operatorname{Var_{boot}}\) (Section~\ref{sec-boot-var}). This dataset
included 1,954 completely forested plots, as well as 95 completely
nonforested plots that were identified as true zeroes (AGB) based on
LiDAR-derived maximum heights \(\leq\) 1 m (L. K. Johnson et al. 2022),
for a total of 2,049 plots inventoried between 2002 and 2019. We used
the map assessment set to estimate the pixel-level residual variances
(Section~\ref{sec-resvarmod}) that contribute to
\(\operatorname{Var_{res}}\) (Section~\ref{sec-res-var}). This dataset
contained 545 plots inventoried in 2007, 2012, 2018, and 2019,
collectively representing a complete spatial coverage of NYS. When plots
in model development or map assessment sets were inventoried more than
once, single instances were randomly selected so that each plot was
associated with only one inventory year. Refer to L. K. Johnson et al.
(2023) for further details on the FIA plot design and the plot selection
criteria used here.

Rather than using the tree-level AGB predictions provided in the FIA
database, we replicated FIA's component-ratio-method (CRM; Woodall et
al. (2011)) to incorporate measurement error, misclassifications, and
allometric uncertainty in the plot-level predictions. At the time of
writing, FIA had transitioned to using a new system of allometric models
(Westfall et al. 2023). However, our use of CRM remained consistent with
the methods used in L. K. Johnson et al. (2023). A flowchart describing
data inputs, component models, and final predictions for CRM is included
in Appendix A.

For every bootstrap iteration we selected a set of plots with
replacement from the model development partition, yielding as many sets
of plots as bootstrap iterations. Further, for each plot and bootstrap
iteration we constructed a plot polygon incorporating location
uncertainty and computed plot-level biomass incorporating both
measurement and classification errors as well as allometric uncertainty
as detailed in the following sub-sections.

\hypertarget{sec-tree-err}{%
\subsubsection{Measurement and classification
error}\label{sec-tree-err}}

Both measurement and classification errors were randomly selected and
assigned to individual trees before summing tree-level predictions for
each plot. We assumed normal distributions, despite representing (at
best) approximations of true error distributions, for measured variables
using parameters (mean, standard deviation) reported in the literature
(Table~\ref{tbl-measurement-err}). DECAYCD misclassification was
randomly assigned by selecting a random number between 0 and 1, and
randomly choosing a sign (+/-). If the selected number was greater than
the estimated 59\% accuracy rate, then the decay class above or below
was assigned based on the randomly selected sign
(Table~\ref{tbl-classification-acc}). Species classifications (SPCD in
FIA DB), percent of bole volume that is represented by bark
(BARK\_VOL\_PCT in FIA DB), and structural loss adjustments (Domke,
Woodall, and Smith 2011) based on assigned decay codes (DECAYCD in FIA
DB) were omitted from tree-level uncertainty quantification due to lack
of data or intractability. For example, misclassification rates for
tree-level species were estimated by Yanai et al. (2023), but a method
for assigning a replacement species in the event of a misclassification
was not apparent.

\hypertarget{sec-allom-err}{%
\subsubsection{Allometric model uncertainty}\label{sec-allom-err}}

We utilized the legacy tree database, which provides records for volumes
and weights of trees and tree components (e.g.~foliage, top and
branches, bole) for North American tree species (Radtke et al. 2015) to
account for allometric model uncertainty. We selected only those records
obtained from sites or studies within FIA's designated Northeastern
states because the CRM models for NYS are specific to this region
(Woodall et al. 2011). We used the database to produce separate hardwood
and softwood empirical residual distributions for bole volume, bole bark
biomass, bole wood biomass, foliage biomass, and total tree biomass
(Appendix B). We did not account for stump biomass uncertainty or error
since stump biomass measurements were not available for the Northeastern
partition of the legacy tree database, and tractable estimates of
uncertainty or error were not presented with the stump biomass models
published in Raile (1982). All biomass residuals were computed relative
to total tree biomass predicted with Jenkins et al. (2003) allometric
models, and volume residuals were computed relative to predicted bole
volume with Scott (1981) models so that adding randomly selected
residuals would produce new estimates that were within a reasonable
distance from the original.

Under the CRM approach, top and branch biomass (TAB) is predicted by
subtracting all other predicted component biomass (stump, bole, foliage)
from the predicted total. To ensure that our random residual selection
yielded TAB estimates that were ecologically sound, we also compiled
ratios of TAB to total tree biomass (minus predicted stump biomass;
Appendix B) using the legacy tree database. We did not partition the TAB
ratio distribution further into softwood and hardwood distributions
since there were limited database records with both TAB measurements and
total tree biomass measurements. Given the paucity of TAB ratio data, we
filtered three records from the TAB ratio distribution where ratios were
\(>\) 0.5 to limit their influence on the overall outcome. After all
components (total biomass, bole biomass, stump biomass, and foliage
biomass) had been predicted and randomly selected residuals had been
added to each component, we randomly selected a TAB ratio, computed a
TAB prediction ({[}TAB ratio{]} \(\cdot\) {[}total biomass{]}), and
summed all components. The difference between the sum of components and
the total was proportionally subtracted from each component to ensure
that the sum of predicted component biomass was equal to the predicted
total tree biomass.

\hypertarget{sec-loc-err}{%
\subsubsection{Plot location uncertainty}\label{sec-loc-err}}

FIA plots are composed of four identical circular subplots with radii of
7.32 m (24 ft), with one subplot centered at the macroplot centroid and
three subplots located 36.6 m (120 ft) away at azimuths of
\(360\degree\), \(120\degree\), and \(240\degree\) (Bechtold and
Patterson 2005). The FIA program averages individual plot locations
collected over multiple repeat visits to improve upon the precision and
accuracy of field-recorded coordinates (Cooke 2000; Hoppus and Lister
2005). We constructed a normal distribution representing plot location
errors by computing the standard deviation of differences between
field-recorded coordinates and the corresponding average coordinates
across all plots in the model development set
(Table~\ref{tbl-measurement-err}). To represent this positional
uncertainty in \(\operatorname{Var_{boot}}\) (specifically within
\(\operatorname{Var_{ref}}\)), for each plot and at each bootstrap
iteration, we randomly selected a distance \(d\) from the constructed
normal distribution and a random azimuth \(a\) between \(0\degree\) and
\(360\degree\) to create new plot centroids and subsequent plot polygons
that were offset from the original plot centroid by a distance \(d\) m
and an angle \(a\degree\).

\hypertarget{tbl-measurement-err}{}
\begin{table}
\caption{\label{tbl-measurement-err}Parameters and sources for estimating tree- and plot-level measurement
error distributions. Where relative standard deviation (SD) is marked as
`T' (true), actual SDs are to be computed relative to the FIA provided
value using the SD fraction in this table
(i.e.~\(\operatorname{actual SD = SD table \cdot FIA value}\)). }\tabularnewline

\centering\begingroup\fontsize{7}{9}\selectfont

\begin{tabular}[t]{l>{\raggedright\arraybackslash}p{8em}>{\raggedleft\arraybackslash}p{3em}>{\raggedleft\arraybackslash}p{3em}>{\raggedright\arraybackslash}p{3em}ll}
\toprule
\multicolumn{1}{c}{FIA DB Name} & \multicolumn{1}{c}{Description} & \multicolumn{1}{c}{Mean} & \multicolumn{1}{c}{SD} & \multicolumn{1}{c}{Rel. SD} & \multicolumn{1}{c}{Units} & \multicolumn{1}{c}{Source}\\
\midrule
DIA & Diameter at breast height & -0.004 & 0.55 & F & cm & Yanai et al. 2023\\
\addlinespace
BOLEHT & Bole height & -0.050 & 1.52 & F & m & Yanai et al. 2023\\
\addlinespace
CULL & Cull, percent rotten or missing & 0.100 & 3.50 & F & \% & Yanai et al. 2023\\
\addlinespace
WOOD\_SPGR\_GREENVOL\_DRYWT & Wood specific gravity (green, oven-dry) & 0.000 & 0.10 & T & NA & Ross 2021\\
\addlinespace
BARK\_SPGR\_GREENVOL\_DRYWT & Bark specific gravity (green, oven-dry) & 0.000 & 0.10 & T & NA & Assumed\\
\addlinespace
STANDING\_DEAD\_DECAY\_RATIO (hardwood) & Hardwood density reduction ratios for DECAYCD 1-5 & 0.000 & 0.01 0.02 0.04 0.05 0.05 & T & \% & Harmon et al. 2011\\
\addlinespace
STANDING\_DEAD\_DECAY\_RATIO (softwood) & Softwood density reduction ratios for DECAYCD 1-5 & 0.000 & 0.01 0.01 0.03 0.03 0.03 & T & \% & Harmon et al. 2011\\
\addlinespace
LON, LAT & Distance between field recorded plot location and plot location averaged over multiple repeat visits & 0.000 & 7.05 & F & m & NYS FIA DB\\
\bottomrule
\end{tabular}
\endgroup{}
\end{table}

\hypertarget{tbl-classification-acc}{}
\begin{table}
\caption{\label{tbl-classification-acc}Classification accuracies. User's accuracy reported for LCMAP. }\tabularnewline

\centering\begingroup\fontsize{8}{10}\selectfont

\begin{tabular}[t]{lllrl}
\toprule
\multicolumn{1}{c}{Name} & \multicolumn{1}{c}{Data Type} & \multicolumn{1}{c}{Description} & \multicolumn{1}{c}{Accuracy (\%)} & \multicolumn{1}{c}{Source}\\
\midrule
DECAYCD & FIA & Tree-level decay code assigned in the field & 59 & Yanai et al. 2023\\
\cmidrule{1-5}
Developed &  &  & 80 & \\

Cropland &  &  & 46 & \\

Grass/shrub &  &  & 37 & \\

Tree cover &  &  & 92 & \\

Wetland &  &  & 97 & \\

Barren & \multirow{-6}{*}{\raggedright\arraybackslash LCMAP} & \multirow{-6}{*}{\raggedright\arraybackslash Primary landcover classification (LCPRI)} & 1 & \multirow{-6}{*}{\raggedright\arraybackslash Stehman et al. 2021}\\
\bottomrule
\end{tabular}
\endgroup{}
\end{table}

\hypertarget{sec-aux-data}{%
\subsection{Auxiliary data}\label{sec-aux-data}}

We used the same set of 29 predictor layers derived in L. K. Johnson et
al. (2023) for model training and mapping. These predictors included
temporally segmented, gap-filled, and smoothed annual Landsat timeseries
imagery and disturbance metrics (Landtrendr; Kennedy, Yang, and Cohen
(2010), Kennedy, Yang, et al. (2018)), LCMAP primary and secondary
landcover classifications, and static ecological, climatic, and
topographic datasets. Each of the 29 predictor layers were projected to
match Landsat 30 m pixel geometries. The raster stacks of predictors
were cropped and aggregated (weighted average) at the constructed FIA
plot polygons (Section~\ref{sec-boot-var}; Section~\ref{sec-loc-err}) to
produce training datasets, and were cropped to the boundaries of the
parcels in our sample for mapping (Section~\ref{sec-studyarea}). The
exactextractr (Baston 2022) and terra (Hijmans 2023) packages for the R
(R Core Team 2023) programming language were used to create the plot-
and parcel-level auxiliary datasets. Further information describing the
auxiliary data can be found in L. K. Johnson et al. (2023).

\hypertarget{sec-agbmodelmap}{%
\subsection{Aboveground biomass models, maps, and
estimates}\label{sec-agbmodelmap}}

Following the `direct' approach described in L. K. Johnson et al.
(2023), ML ensemble models were fit to the resampled plots and
corresponding AGB predictions for each bootstrap iteration
(Section~\ref{sec-boot-var}). Each ensemble was a linear regression that
combined predictions from a random forest implemented with the ranger R
package (Breiman 2001a; Wright and Ziegler 2017), a stochastic gradient
boosting machine as implemented in the lightgbm R package (Friedman
2002; Ke et al. 2017; Shi et al. 2022), and a support vector machine
(SVM) as implemented in the kernlab R package (Cortes and Vapnik 1995;
Karatzoglou et al. 2004). We used the same set of hyperparameters,
identified in L. K. Johnson et al. (2023) as the most performant via an
iterative grid search, for each respective ML model across all bootstrap
iterations. We used these pre-selected hyperparameters not only because
the iterative grid search used in L. K. Johnson et al. (2023) was
computationally expensive, but also because we do not expect the AGB or
predictor distributions within each bootstrap sample to fundamentally
differ since they represent the same underlying population. Further
information describing the ML ensembles and hyperparameter tuning can be
found in L. K. Johnson et al. (2023).

To compute \(\operatorname{Var_{boot}}\), an estimate of mean AGB for
the parcel was computed for each iteration of the bootstrap
(Section~\ref{sec-boot-var}; Equation~\ref{eq-varboot}). To this end,
separate sets of AGB predictions were made for all pixels within each
sampled parcel using the models trained for each bootstrap iteration.
Further, a separate forest mask was created and applied to each parcel
for each bootstrap iteration to incorporate uncertainty and error in the
LCMAP primary landcover classifications. New forest masks were produced
at each iteration by generating a uniform random number between 0 and 1
for each pixel. If this number was greater than the user's accuracy rate
for the pixel's initial classification (LCPRI), then the LCMAP secondary
classification (LCSEC) was used instead
(Table~\ref{tbl-classification-acc}). After the new mask was generated,
the mask was used to identify pixels that were not classified as tree
cover, wetlands, or grass/shrub (forest definition from L. K. Johnson et
al. (2023)) and set their AGB predictions to 0. After masking, and for
each bootstrap iteration, parcel level means
(Equation~\ref{eq-bootmean}) were computed.

\hypertarget{sec-resvarmod}{%
\subsection{Residual variance model}\label{sec-resvarmod}}

Pixel-level residual variance was required to estimate
\(\operatorname{Var_{res}}\) (Section~\ref{sec-res-var};
Equation~\ref{eq-varres}). However, residuals were only computed for a
sample of pixels coinciding with FIA plots selected for the assessment
partition from L. K. Johnson et al. (2023)
(Section~\ref{sec-field-data}). Further, assigning the residual variance
computed from this sample to every pixel would incorrectly assume
homoscedasticity. To overcome this, we fit a model of estimated residual
variance as a function of predicted AGB. Following the approach outlined
in McRoberts et al. (2022), we arranged our assessment plots in order of
increasing AGB prediction and partitioned the data into thirty equal AGB
intervals ranging from 0 to the maximum AGB prediction in the dataset.
To ensure a sufficient number of observations were available to estimate
residual variance for each interval (McRoberts et al. 2022), we
collapsed the initial groups until each interval contained at least 10
plots, resulting in 21 groups. We then computed the average AGB
prediction and the residual variance for each group, and fit several
forms of regression models relating AGB to residual variance for the 21
points in the dataset. Specifically, we evaluated linear, log-linear,
log-log, third order polynomial, and natural cubic spline (2 knots,
Hastie (1992)) model forms. The most basic (linear) model form was
constructed as follows:

\begin{equation}\protect\hypertarget{eq-resvarmod}{}{
\hat{\sigma_{i}}^2 = \beta_0 + \beta_1 \cdot \hat{y_i} + \varepsilon
}\label{eq-resvarmod}\end{equation}

where \(\hat{y_i}\) is the average AGB prediction for the group,
\(\hat{\sigma_i}^2\) is the estimated residual variance for the group,
and \(\beta_{0}\) and \(\beta_{1}\) are coefficients estimated via
ordinary least squares. All other model forms were fit similarly, using
the same independent and dependent variables, but with
log-transformations or additional model terms. When log transformations
were used, we back-transformed estimates to a linear scale and applied
multiplicative correction factors defined in Sprugel (1983) to eliminate
log-transformation bias (Baskerville 1972) as follows:

\begin{equation}\protect\hypertarget{eq-sprugel}{}{
\operatorname{CF} = e^{\sigma^{2} / 2}
}\label{eq-sprugel}\end{equation}

where \(\operatorname{sigma}\) is the residual standard deviation from
the fitted model.We chose the model that best fit the data as indicated
by the coefficient of determination (\(\operatorname{R^2}\); computed
after backtransformation and bias-correction), and then used this final
model to estimate the residual variance for each pixel in our parcel
sample.

\hypertarget{sec-spatcormeth}{%
\subsection{Spatial autocorrelation of
residuals}\label{sec-spatcormeth}}

An estimate of spatial autocorrelation between model residuals was
required to estimate \(\operatorname{Var_{res}}\)
(Section~\ref{sec-res-var}; Equation~\ref{eq-varres}). However, the
sample of assessment FIA plots used in L. K. Johnson et al. (2023), with
at most one FIA plot per 2,400 ha in NYS (Bechtold and Patterson 2005),
was too sparse to provide the necessary information. We instead used
LiDAR-based predictions of AGB developed in L. K. Johnson et al. (2022)
as an alternative form of reference data. We acknowledge that these
predictions are of lesser quality than field-measured data, but they
offered improved accuracy relative to the Landsat-based models assessed
here, and they provided the spatial density and coverage that can
represent local patterns of variability in forests. To our knowledge,
these two important qualities are not offered by any other AGB dataset
within NYS. We temporally and spatially matched Landsat-based maps with
the spatiotemporal patchwork of LiDAR-based AGB predictions (L. K.
Johnson et al. 2022) and selected a simple random sample of 500,000
pixels for which residuals were computed (Equation~\ref{eq-residuals}).
We first calculated an empirical semivariogram from our sampled
residuals with a bin width of 30 m (pixel resolution) which we then used
to fit exponential, spherical, and Gaussian model semivariograms. We
assessed the fits of each model semivariogram visually, and selected the
best fitting model to use in Equation~\ref{eq-varrho}. All variogram
computation and fitting was done with the gstat R package (Pebesma 2004;
Gräler, Pebesma, and Heuvelink 2016).

To assess the sensitivity of our uncertainty estimates relative to the
estimate of spatial autocorrelation between our LiDAR-based residuals,
we computed four additional estimates of \(\operatorname{Var_{res}}\)
(Equation~\ref{eq-varres}) where we multiplied pixel pairwise distances
by 0.25, 0.5, 2, and by the estimated range of the semivariogram. In
effect, these additional estimates of \(\operatorname{Var_{res}}\)
reflect residual spatial autocorrelation multiplied by a factor of 4, 2,
1/2, and 0 respectively.

\hypertarget{sec-parcelmod}{%
\subsection{Modeling estimated standard error (total variance) as a
function of parcel characteristics}\label{sec-parcelmod}}

To increase the computational efficiency and practical utility of
estimating spatially aggregated uncertainty for parcels or other
subregions of interest, we modeled parcel SE as a function of parcel
characteristics that we expect are readily available to map users.
Specifically, we included AGB density, \% forest cover, area, and
perimeter of each parcel as predictors of a parcel's SE given that these
predictors are straightforward to compute from an area of interest
polygon, the AGB prediction raster, and a co-located forest mask raster.
Forest cover was defined as the combination of the tree cover, wetlands,
and grass/shrub LCMAP classes (L. K. Johnson et al. 2023), and AGB
density (\(\operatorname{Mg\ ha^{-1}}\)) was computed as the average of
pixel predictions within a parcel (Equation~\ref{eq-bootmean}).

We fit four candidate models, one with no variable transformation, a
second with natural log transformations for both the dependent and
independent variables, and third and fourth models with natural log
transformations for the dependent and independent variables
respectively. When log transformations were used, we back-transformed
estimates to a linear scale and applied correction factors
(Equation~\ref{eq-sprugel}) defined in Sprugel (1983) to eliminate
log-transformation bias (Baskerville 1972). Candidate models were fit as
follows:

\begin{equation}\protect\hypertarget{eq-logmod}{}{
\operatorname{SE} = \beta_0 + \beta_1 \cdot AGB + \beta_{2} \cdot \% Forest + \beta_{3} \cdot Area + \beta_4 \cdot Perimeter + \varepsilon
}\label{eq-logmod}\end{equation}

where \(AGB\), \(\%\ Forest\), \(Area\), and \(Perimeter\) are described
above and \(\beta_{0}\), \ldots, \(\beta_{4}\) are coefficients
estimated via ordinary least squares. Each of the candidate models were
fit to a randomly selected 80\% of the parcel sample (training set) and
were compared on the basis of R\textsuperscript{2}. The candidate model
with the best R\textsuperscript{2} statistic was selected as the final
model. Using the final model and the remaining 20\% of the parcel sample
(testing set) we computed model accuracy metrics including root mean
squared error (RMSE), mean absolute error (MAE), mean error (ME), and
R\textsuperscript{2} using the yardstick R package (Kuhn, Vaughan, and
Hvitfeldt 2023). All model accuracy and comparison metrics were computed
after any necessary backtransformation and bias-correction had been
applied.

\hypertarget{results}{%
\section{Results}\label{results}}

\hypertarget{reference-data-uncertainty}{%
\subsection{Reference data
uncertainty}\label{reference-data-uncertainty}}

Comparing the variability of tree- and plot-level AGB predictions
provides insight into how allometric uncertainty and measurement error
scales from individual trees to aggregations of many individuals within
a plot. Over the 1000 bootstrap iterations tree-level AGB predictions
varied more than plot-level AGB predictions (Appendix C). The average
tree-level AGB standard deviation (SD) and coefficient of variation (CV)
were 65.93 kg and 35.01\%, while the average plot-level AGB SD and CV
were 7.47 \(\operatorname{Mg\ ha^{-1}}\) and 7.20\%. This result was
unsurprising because plot-level predictions are simply area-normalized
sums of tree-level predictions, and aggregating individuals can smooth
out extreme individual variability as predictions above and below
respective means will offset to an extent. Softwoods and hardwood trees
varied similarly in absolute terms (SD), but softwoods varied
substantially more in relative terms (CV).

\hypertarget{spatial-autocorrelation-of-residuals}{%
\subsection{Spatial autocorrelation of
residuals}\label{spatial-autocorrelation-of-residuals}}

The exponential model semivariogram demonstrated the best fit to our
empirical semivariogram (Section~\ref{sec-spatcormeth}; Appendix D) and
was selected to provide estimates of spatial correlation between model
residuals in our calculation of \(\operatorname{Var_{res}}\)
(Equation~\ref{eq-varres}). Residual semivariance increased from 199
\(\operatorname{(Mg\ ha^{-1})^2}\) for neighboring pixels (nugget), to
1253 \(\operatorname{(Mg\ ha^{-1})^2}\) (sill) for pixels roughly 1025 m
(range) apart. Residual autocorrelation stops increasing beyond the
estimated range and thus pixel pairs separated by distances greater than
this range yielded \(\hat{\rho_{ij}}\) = 0 (Equation~\ref{eq-varrho}).

\hypertarget{sec-pixelvar}{%
\subsection{Pixel-level uncertainty}\label{sec-pixelvar}}

\begin{figure}

{\centering \includegraphics{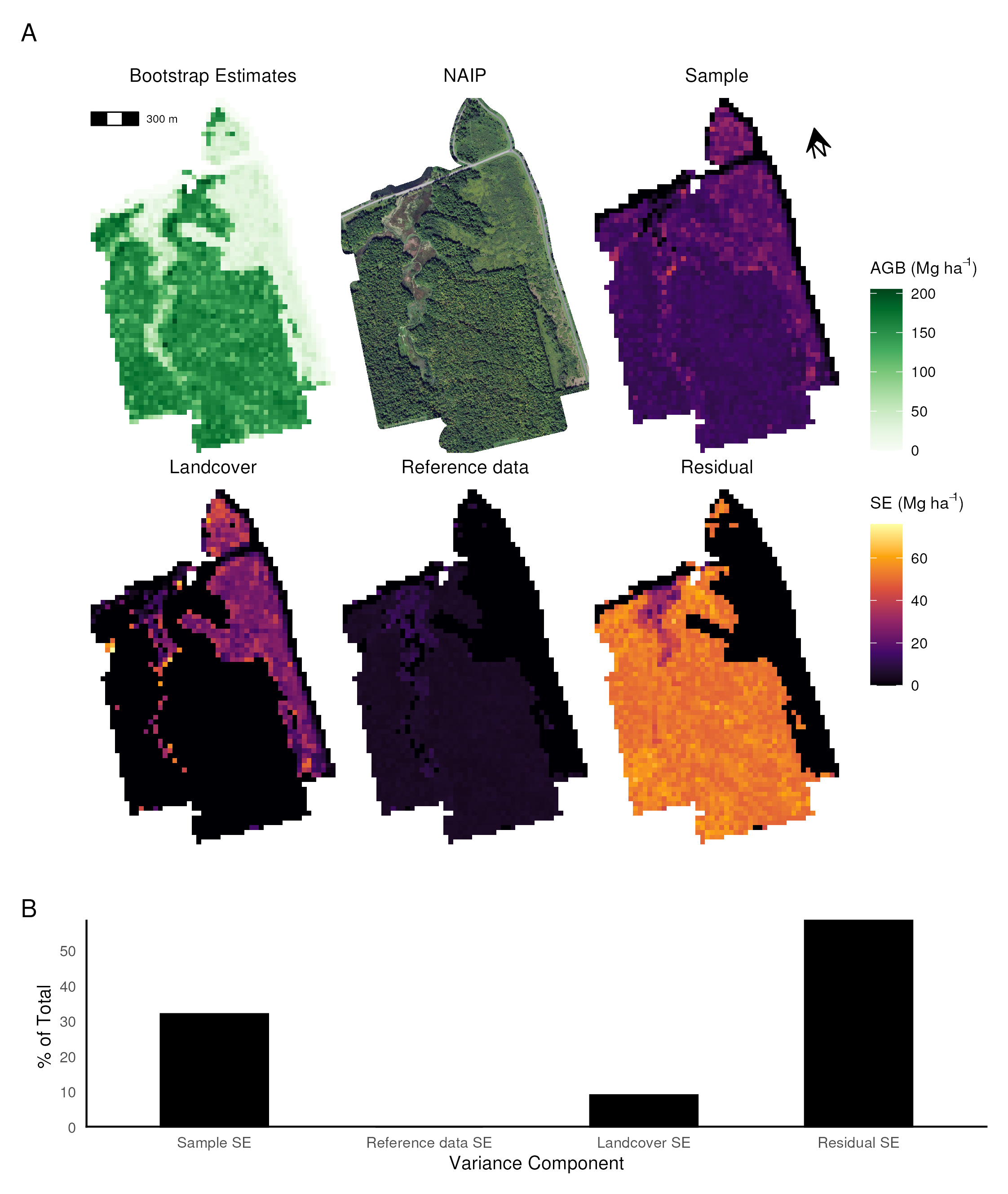}

}

\caption{\label{fig-parcelviz}Aggregating pixel-level uncertainties to
produce parcel-level standard error (SE). A) Bootstrap estimates: AGB
estimates computed as the mean prediction over all bootstrap iterations.
National Aerial Imagery Program (NAIP) orthophotography collected in
2019 for additional reference. Sampling standard error (SE), landcover
SE, reference SE, and residual SE estimates. B) Relative contributions
(\%) to total SE for the area of interest in A.}

\end{figure}

The natural cubic spline model (Section~\ref{sec-resvarmod}) fit the
grouped data best and explained 55\% of the variability in residual
variance (coefficient of determination; \(\operatorname{R^2}\); Appendix
E). Maps of pixel-level AGB estimates and co-located residual SEs
(Figure~\ref{fig-parcelviz} A) demonstrated the modeled relationship
spatially, with regions of greater AGB density yielding larger residual
SE estimates. Conversely, the largest sample and landcover SEs
co-occurred within areas of consistently small AGB estimates
(Figure~\ref{fig-parcelviz} A). This sample SE pattern may be explained
by the relative lack of model training data at the lower extreme of the
distribution (L. K. Johnson et al. 2023) while the landcover SE pattern
may be explained by the inclusion of LCMAP classification uncertainty in
our bootstrapping procedure (Section~\ref{sec-agbmodelmap}). Marginal or
transitional lands with relatively low AGB density are among the most
challenging to accurately classify (Grass/Shrub and Cropland;
Table~\ref{tbl-classification-acc}) and may be intermittently masked to
0 AGB across bootstrap iterations. Pixel-level reference data SE was
consistently near zero.

\hypertarget{spatially-aggregated-uncertainty-estimates}{%
\subsection{Spatially aggregated uncertainty
estimates}\label{spatially-aggregated-uncertainty-estimates}}

\begin{figure}

{\centering \includegraphics{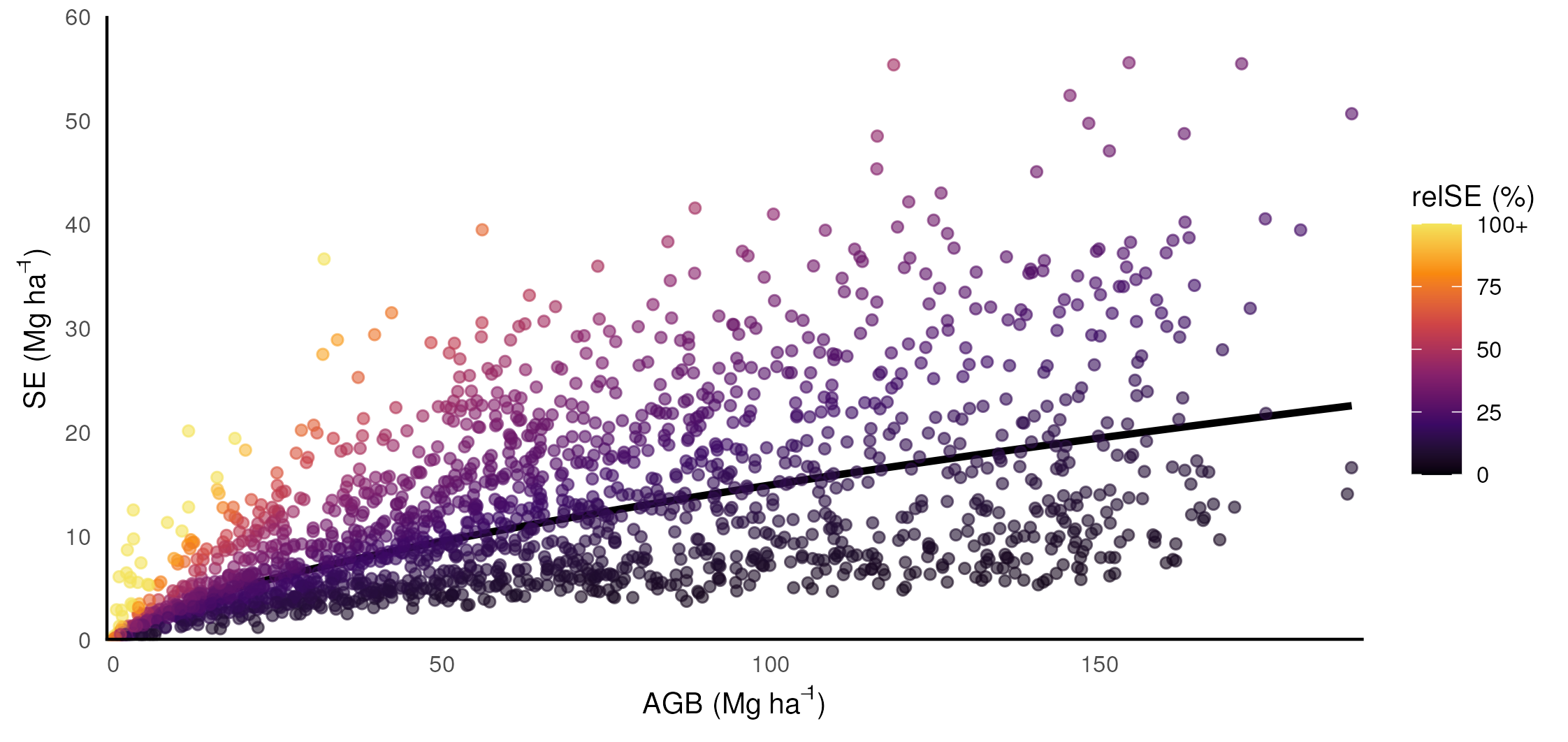}

}

\caption{\label{fig-relse}Parcel-level standard error (SE) as a function
of AGB density. Each point represents a parcel from the training
partition of the parcel sample, and each point is shaded by the relative
standard error (relSE; \(\operatorname{(SE\ /\ AGB) \cdot 100}\)).
Parcels with relSE \textgreater{} 100\% were set to 100\% for
visualization. The black trend line was fit to data using a natural
log-log regression.}

\end{figure}

\begin{figure}

{\centering \includegraphics{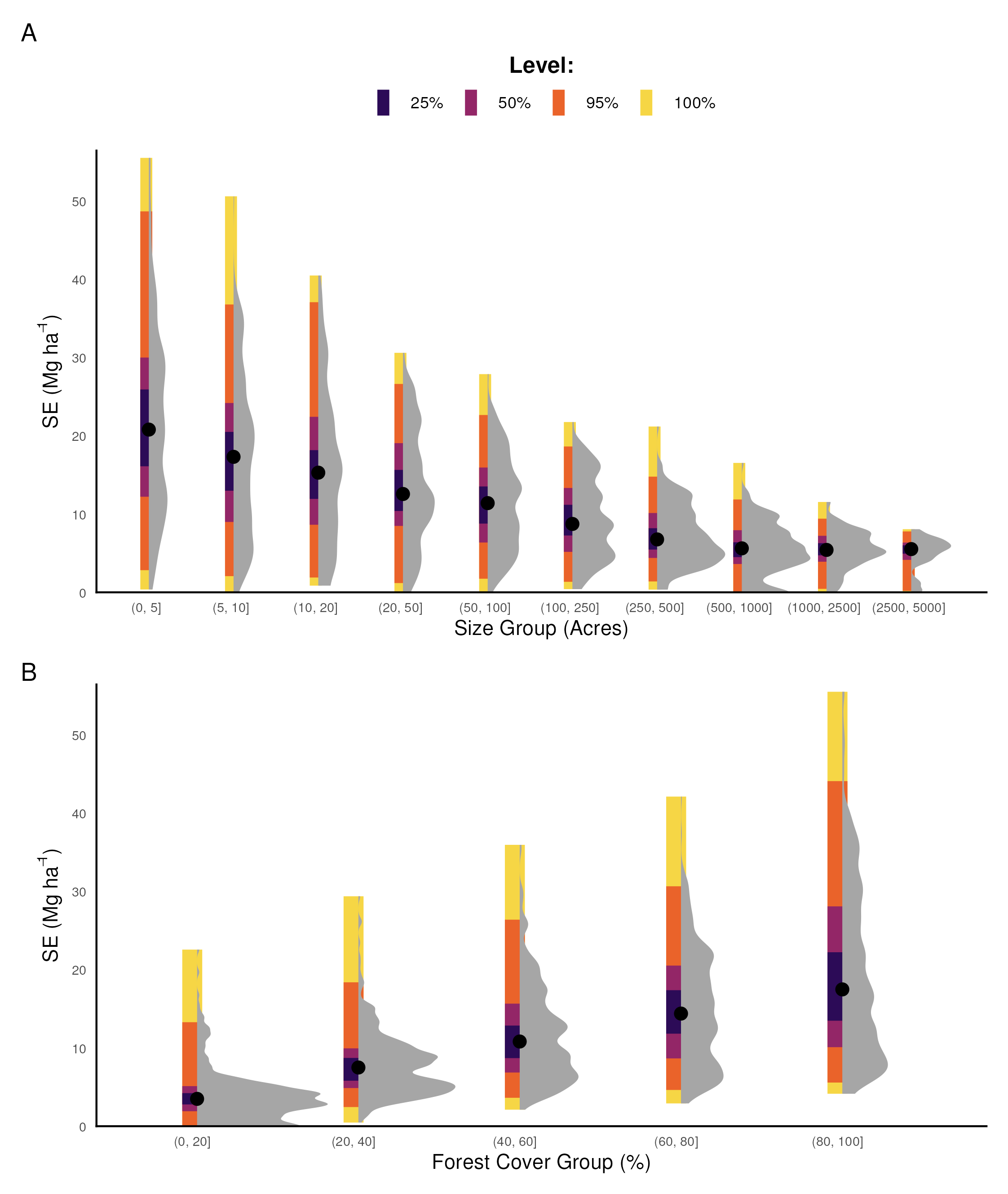}

}

\caption{\label{fig-groupdist}Parcel-level standard error (SE)
distributions. Gray shaded areas represent smoothed frequency
distributions of SEs, black dots identify median values, and colored
bars show 25\%, 50\%, 95\%, and 100\% intervals (percentiles). A)
Distributions per size group (acre). B) Distributions per forest cover
group (\%). All parcels in the training partition of the parcel sample
are included.}

\end{figure}

Parcel-level estimates of standard error (SE; Equation~\ref{eq-se})
ranged from near 0 \(\operatorname{Mg\ ha^{-1}}\) to \textasciitilde60
\(\operatorname{Mg\ ha^{-1}}\) but varied substantially with parcel
size, forest cover, and AGB density (Figure~\ref{fig-relse};
Figure~\ref{fig-groupdist}). Although the range of SE was large, the
majority of relative standard errors (relSE =
\(\operatorname{SE\ /\ AGB \cdot 100}\)) were \(\leq 25\%\) and only
exceeded 50\% for \textasciitilde7\% of parcels in the training
partition (Figure~\ref{fig-relse}). Parcels with the smallest AGB
density yielded the largest relSEs.

\begin{figure}

{\centering \includegraphics{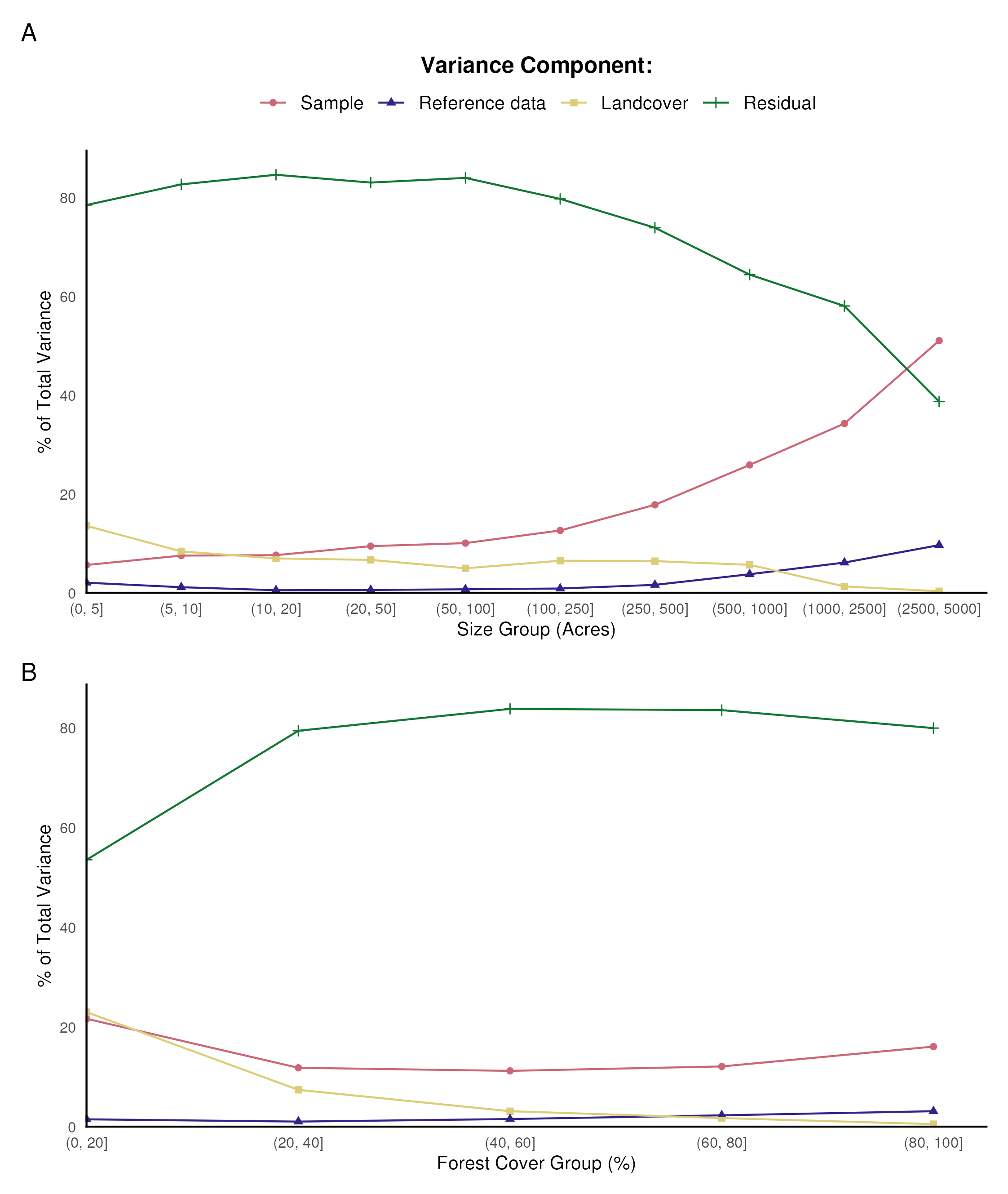}

}

\caption{\label{fig-contrib}Uncertainty contributions (\% of total
variance) as a function of parcel size (A) and forest cover (B) for the
training partition of the parcel sample. For each source of uncertainty,
the average percent of the total variance was summarized across all
parcels within each group.}

\end{figure}

Given that the formula for \(\operatorname{Var_{res}}\)
(Equation~\ref{eq-varres}) contained \(\operatorname{N^2}\) (N = number
of pixels) in the denominator, it is intuitive that
\(\operatorname{Var_{total}}\) (Equation~\ref{eq-vartot}) and SE
(Equation~\ref{eq-se}) decreased as a function of parcel size; the
larger a parcel, the more 30 m pixels contained therein. Further, as
parcel size grew beyond the range of spatial correlation and average
pixel distances increased, it follows that \(\operatorname{Var_{res}}\)
decreased accordingly, and we might expect that
\(\operatorname{Var_{res}}\) will eventually become negligible for
parcels larger than those considered in this study
(Figure~\ref{fig-contrib}). Forest cover impacted parcel SE in the
opposite direction (Figure~\ref{fig-groupdist}), where increasing forest
cover resulted in increased standard error. Non-forest pixels resulted
in both 0 AGB and 0 residual variance; thus the greater number of these
pixels in a parcel, the less forest cover and the smaller the
contributions of \(\operatorname{Var_{res}}\) to
\(\operatorname{Var_{total}}\). However, the relationship between forest
cover and SE may have been somewhat weakened by the impact of LCMAP
classification uncertainty (\(\operatorname{Var_{LC}}\)), where the
relatively poor accuracy of cropland and grass/shrub classified pixels
may have yielded greater bootstrap variances as pixels were
intermittently masked to 0 AGB across bootstrap iterations
(Table~\ref{tbl-classification-acc}; Section~\ref{sec-agbmodelmap};
Section~\ref{sec-pixelvar}).

\(\operatorname{Var_{res}}\) was greater than all other variance
components across most size and forest cover groups
(Figure~\ref{fig-contrib}). The contribution of
\(\operatorname{Var_{sam}}\) was only larger than
\(\operatorname{Var_{res}}\) for the largest parcels (2500 to 5000
acres) where spatial autocorrelation of residuals had a diminished
impact. \(\operatorname{Var_{LC}}\) contributed substantially (\(\geq\)
20\%) for the least forested parcels, but contributed \(\leq\) 10\% for
all but the smallest forest parcels.\(\operatorname{Var_{ref}}\)
contributed the least of any other component for all but the largest and
most forested parcels (1000-5000 acres; 80-100\% forested), and was
always \(< 10\%\) of \(\operatorname{Var_{total}}\).

\begin{figure}

{\centering \includegraphics{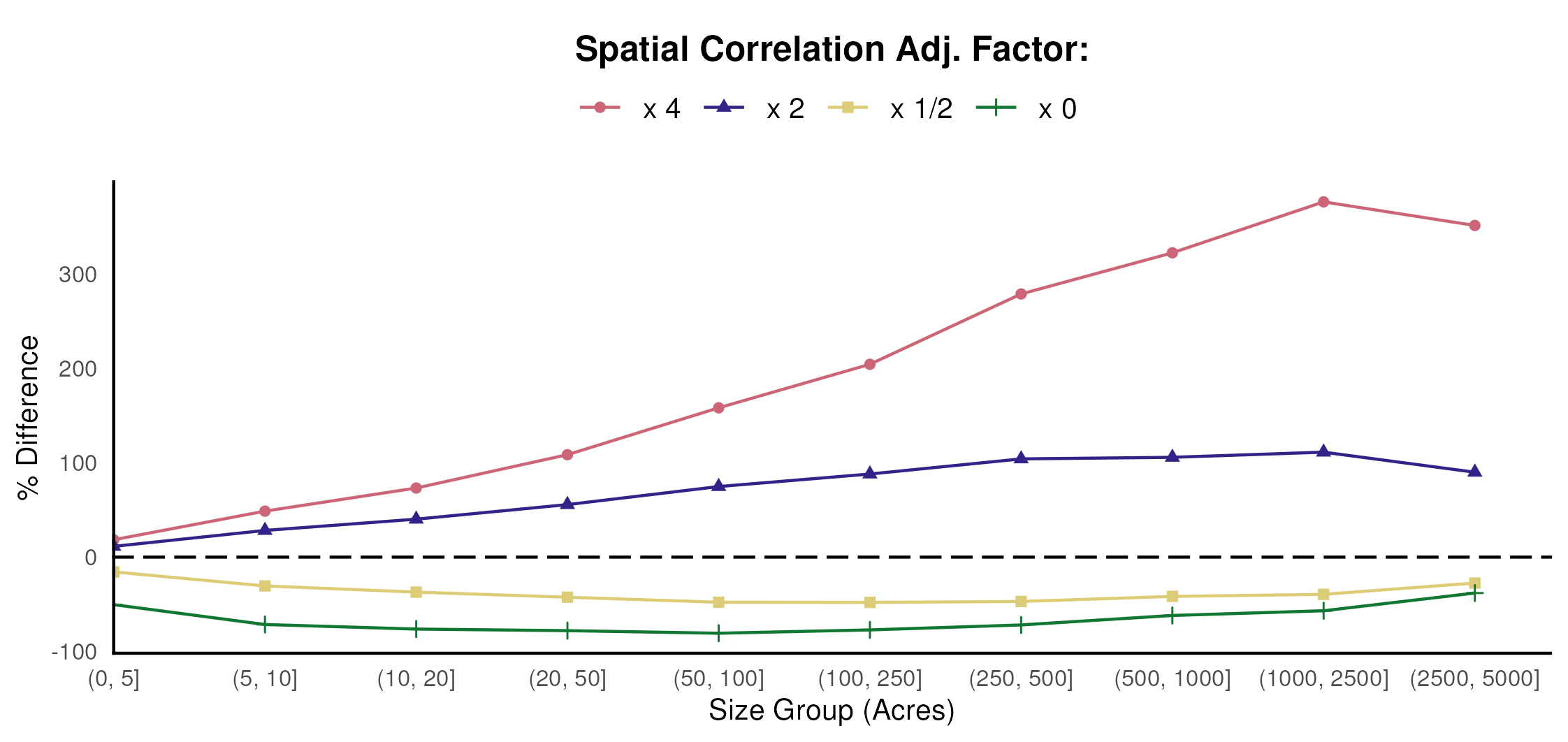}

}

\caption{\label{fig-spatcorr-sens}Sensitivity of total variance
estimates to strength of residual spatial autocorrelation. Percent
difference between total variance adjusted for strengthened/weakened
spatial correlation of residuals (adjustment factor \(\neq\) 1) and
total variance computed regularly (adjustment factor = 1; dashed black
line) as a function of parcel size (size group) for the training
partition of the parcel sample. The average percent difference was
summarized across all parcels within each group. Spatial correlation
adjustment factors are used to modify the pairwise pixel distances
(e.g.~adjustment factor 4 is implemented by dividing the pairwise pixel
residual distances by 4) such that the adjustment factor effectively
describes a multiplicative adjustment to the empirical spatial
correlation of residuals.}

\end{figure}

Parcel-level estimates of \(\operatorname{Var_{total}}\) (and
subsequently SE) were sensitive to the strength of residual spatial
autocorrelation (Section~\ref{sec-spatcormeth};
Figure~\ref{fig-spatcorr-sens}). Adjustment factors that strengthened
residual autocorrelation had the smallest impact for small parcels,
where the majority of pair-wise pixel distances were short and residual
spatial autocorrelation already had a near-maximum impact. On the other
hand, strengthening residual autocorrelation increased
\(\operatorname{Var_{total}}\) by \(\geq\) 50\% for all size groups
\(\geq\) 20 acres. Halving the strength of residual spatial correlation
reduced \(\operatorname{Var_{total}}\) from 15.7\% to 48.0\%, and
removing residual spatial correlation altogether reduced
\(\operatorname{Var_{total}}\) from 38.1\% to 80.7\%. This highlights
the importance of residual spatial autocorrelation as a source of
uncertainty when estimating SE for small areas.

\newpage

\hypertarget{modeling-estimated-standard-error-total-variance-as-a-function-of-parcel-characteristics}{%
\subsection{Modeling estimated standard error (total variance) as a
function of parcel
characteristics}\label{modeling-estimated-standard-error-total-variance-as-a-function-of-parcel-characteristics}}

\begin{figure}

{\centering \includegraphics{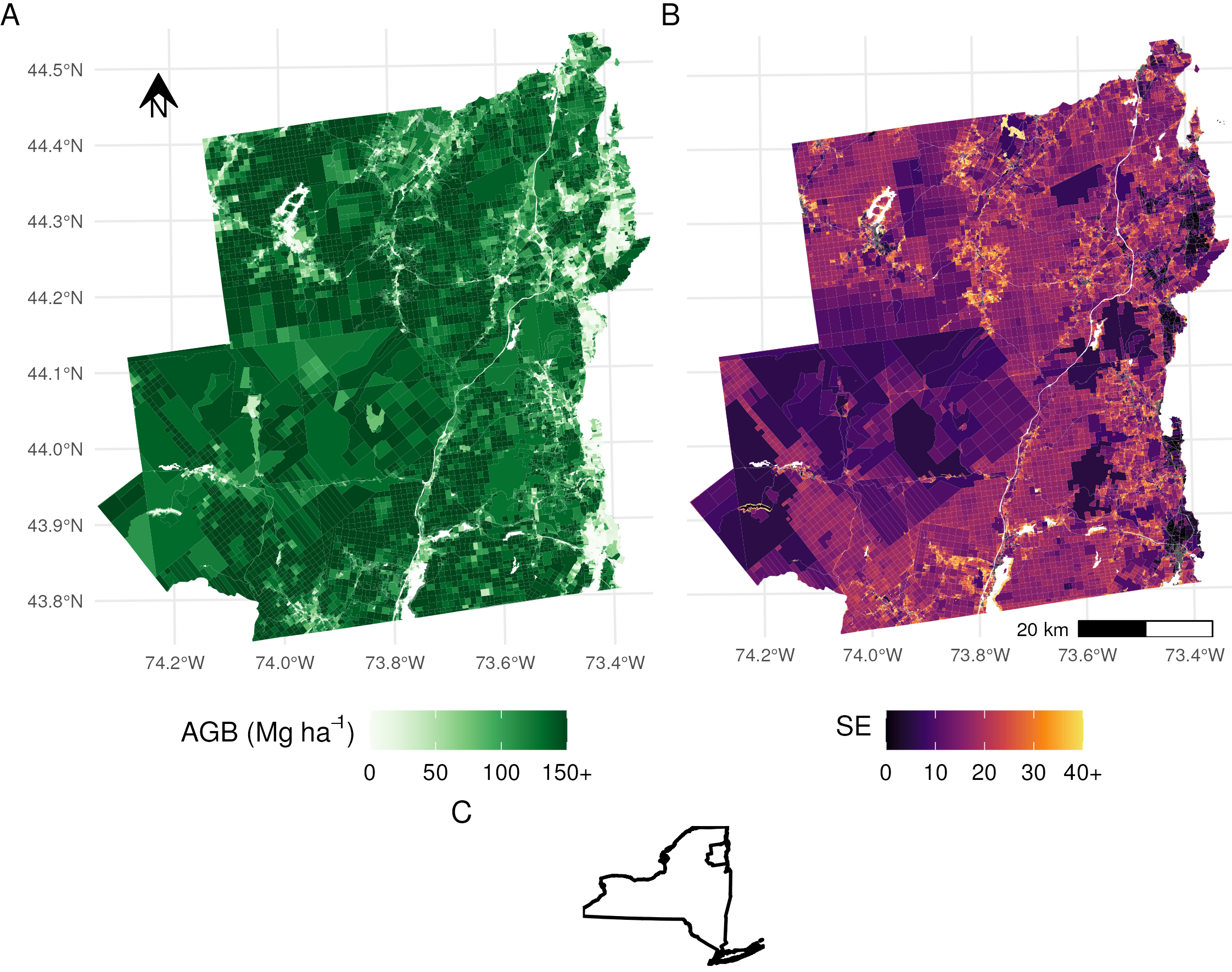}

}

\caption{\label{fig-essex-cty}Estimates of aboveground biomass density
and associated standard error for ownership parcels in Essex County,
NYS. A) Parcel-level AGB density computed as the mean of
pixel-predictions within each parcel. B) Parcel-level SE estimated from
a natural log-log regression with parcel area, perimeter, AGB density,
and \% forest cover as independent variables. C) Location of Essex
County within NYS. AGB density and SE values truncated at 150 and 40
respectively for display.}

\end{figure}

The natural log-log regression achieved the best R\textsuperscript{2}
value among the candidate models (Section~\ref{sec-parcelmod}) and
accurately estimated SEs for parcels in the test set (Appendix F; RMSE =
2.46, MAE = 1.38, ME = -0.21, R\textsuperscript{2} = 0.93). The
estimated coefficients (Table~\ref{tbl-se-lm}) indicate that AGB density
was the most important predictor of SE, which follows the relationship
inherent to our residual variance model (Section~\ref{sec-resvarmod};
Appendix E) and is corroborated by both pixel-level results
(Figure~\ref{fig-parcelviz}) and relSE results (Figure~\ref{fig-relse}).
Parcel area, perimeter, and \% forest cover were similarly important
predictors in the model. Using this model we estimated parcel-level SE
for all ownership parcels in Essex County, New York State (NYS;
Figure~\ref{fig-essex-cty}).

\hypertarget{tbl-se-lm}{}
\begin{table}
\caption{\label{tbl-se-lm}Results of a natural log-log regression model of estimated parcel-level
standard error as a function of parcel characteristics (AGB, Perimeter,
Area, and \% Forest). }\tabularnewline

\centering
\begin{tabular}[t]{lrrrr}
\toprule
\multicolumn{1}{c}{ } & \multicolumn{1}{c}{Coefficient} & \multicolumn{1}{c}{Standard error} & \multicolumn{1}{c}{t} & \multicolumn{1}{c}{p}\\
\midrule
Intercept & 1.298 & 0.088 & 14.8 & < 0.001\\
\addlinespace
ln(AGB) & 0.584 & 0.006 & 105.4 & < 0.001\\
\addlinespace
ln(Perimeter) & -0.167 & 0.014 & -12.0 & < 0.001\\
\addlinespace
ln(Area) & -0.146 & 0.008 & -17.9 & < 0.001\\
\addlinespace
ln(\% Forest) & 0.123 & 0.005 & 26.7 & < 0.001\\
\bottomrule
\end{tabular}
\end{table}

\hypertarget{discussion}{%
\section{Discussion}\label{discussion}}

In this study we produced model-based estimates of uncertainty (standard
error; SE) for small-area (\textless1-5000 acres) spatial averages of
forest aboveground biomass (AGB) maps. Our SE estimation approach
uniquely included all four types of uncertainty
(Table~\ref{tbl-uncertainty-types}), was compatible with algorithmic
models (machine learning or nonparametric), and incorporated residual
spatial autocorrelation. We found that residual variance
(\(\operatorname{Var_{res}}\); Equation~\ref{eq-varres}) dominated other
variance components for all but the largest parcels (2500-5000 acres),
owing largely to contributions from residual spatial autocorrelation
(Figure~\ref{fig-spatcorr-sens}). These results underscore the necessity
of incorporating residual spatial autocorrelation when estimating
uncertainty at scales germane to forest management (e.g.~parcel sizes in
New York state). We demonstrated that a log-log regression model for
estimating SE as a function of parcel characteristics was very accurate,
opening the door to more computationally efficient and practical
uncertainty estimation for map subregions. Overall, these findings pave
the way for rigorous uncertainty estimation in future regional-scale
model-based forest carbon accounting and monitoring efforts.

\hypertarget{sec-contrib-disc}{%
\subsection{Contributions to total uncertainty}\label{sec-contrib-disc}}

Attributing contributions to total uncertainty from various sources is a
critical step to reducing uncertainty, one of the IPCC good practice
guidelines (``uncertainties are reduced as far as practicable''; Buendia
et al. (2019)). Without identifying sources of uncertainty, reduction
efforts cannot be effectively directed. Although these results most
directly inform reducing uncertainties associated with the specific
modeling framework implemented in L. K. Johnson et al. (2023), the
general patterns uncovered here may also serve to inform future
model-based efforts insofar as sufficient overlap exists in terms of
sampling designs, reference data collection protocols, remotely sensed
data products, and model forms.

Residual variance (\(\operatorname{Var_{res}}\);
Equation~\ref{eq-varres}) dominated all sources of uncertainty for all
but the largest parcel sizes (0-2500 acres; Figure~\ref{fig-contrib})
primarily due to the impact of model residual spatial autocorrelation
(Figure~\ref{fig-spatcorr-sens}). These results hold significance for
future model-based efforts in areas where forests are divided into
smaller units and parcel dimensions do not far exceed the range of
spatial autocorrelation. In regions like the eastern and midwestern
United States (US), forests are predominantly managed by nonindustrial
private forest owners (``family forests''), and these family forest
holdings are often small (Nelson, Liknes, and Butler 2010; Butler et al.
2016). We therefore recommend prioritizing model accuracy to most
effectively reduce total uncertainty (\(\operatorname{Var_{total}}\),
Equation~\ref{eq-vartot}; SE, Equation~\ref{eq-se}) in these contexts.
Increased model accuracy simultaneously reduces both residual variance
and residual spatial autocorrelation, effectively diminishing
\(\operatorname{Var_{res}}\). Of course, model accuracy is almost always
a priority, but we highlight its importance relative to other
uncertainty components assuming a limited amount of time and resources
with which to reduce total variance (Table~\ref{tbl-uncertainty-types}).
In particular, the inclusion of data from active remote sensing systems,
like synthetic aperture radar or spaceborne LiDAR which can penetrate
forest canopies and may better represent structural attributes related
to AGB, could increase model accuracy (Quegan et al. 2019; Dubayah et
al. 2020). Higher resolution imagery in concert with the implementation
of deep-learning algorithms like convolutional neural networks which can
exploit local spatial patterns may offer improved model predictions as
well (LeCun, Bengio, and Hinton 2015; Kattenborn et al. 2021).

Sampling variance (\(\operatorname{Var_{sam}}\);
Equation~\ref{eq-varsam}) was the largest contributor to total
uncertainty for only the largest parcels (2500-5000 acres;
Figure~\ref{fig-contrib}). We expect this to hold true for NYS parcels
larger than the maximum size considered in this study given the inverse
relationship between \(\operatorname{Var_{res}}\) and parcel size. In
other words, while we assume \(\operatorname{Var_{res}}\) will approach
zero with increasing parcel size, we also assume that
\(\operatorname{Var_{sam}}\) will remain relatively stable and thus
contribute a larger and larger proportion to a decreasing
\(\operatorname{Var_{total}}\). With all other variables held constant,
this suggests that in regions where forests are contained in larger
contiguous management units, like the western US (Nelson, Liknes, and
Butler 2010; Butler et al. 2016), reductions in model-based uncertainty
can be most efficiently achieved by investing in more forest inventory
sample locations. However, forests in the western US are generally
conifer-dominated with even-aged disturbance regimes (Perry et al. 2022;
Rogers 1996), whereas forests in the eastern US are mostly mixed
hardwood-conifer with uneven-aged disturbance regimes (Seymour, White,
and deMaynadier 2002; Dyer 2006). As a result, it is entirely possible
that spatial correlation of model residuals may be stronger across the
larger and more uniform patches of forest in the West, suggesting that
reducing residual variance (\(\operatorname{Var_{res}}\)) instead of
sampling variance (\(\operatorname{Var_{sam}}\)) may be the most
efficient means to reduce total variance. That said, eastern and western
forests of the US may be too distinct to be able to extrapolate drivers
of model-based variability from one region to the other.

Contributions from landcover mask variability
(\(\operatorname{Var_{LC}}\); Equation~\ref{eq-varlc}) were small except
in the least forested parcels (0-20\% forested; \textgreater{} 20\% of
\(\operatorname{Var_{total}}\); Figure~\ref{fig-contrib}). We can infer
that the increased landcover variability in these parcels is tied to the
presence of marginal and transitional landcover classes
(e.g.~grass/shrub, cropland), which in many contexts may hold
non-trivial quantities of forest biomass (Schnell et al. 2014; K. D.
Johnson et al. 2015; S. Liu et al. 2023), but are traditionally
challenging to classify (Table~\ref{tbl-classification-acc}; Brown et
al. (2020), Stehman et al. (2021), Mahoney et al. (2022)). Subsequently,
we suggest that landcover classification accuracy and precision may be
worth prioritizing in areas dominated by woodlands, shrublands, other
marginal, transitional, or novel land cover types, or where agroforestry
is commonplace (Van Auken 2000; Knapp et al. 2007; Schoeneberger 2009;
Udawatta and Jose 2011; Mahoney et al. 2022).

Reference data variability (\(\operatorname{Var_{ref}}\);
Equation~\ref{eq-varref}) contributed minimally to total uncertainty
across all analyzed groupings (Figure~\ref{fig-contrib}; \textless{}
10\% of \(\operatorname{Var_{total}}\)). These results corroborate
previous studies reporting that tree-level uncertainty contributes
minimally to total uncertainty when aggregated over many individuals
(Breidenbach et al. 2014; Ståhl et al. 2014; McRoberts and Westfall
2014; Yanai et al. 2023); indeed, we demonstrated that the coefficient
of variation (CV) for plot-level AGB was an order of magnitude smaller
than the CV for tree-level AGB (tree-level AGB CV 35.01\%; plot-level
AGB CV 7.2\%) despite relying on relatively small-area plots (0.067 ha).
One would expect even less variability if larger area plots were
utilized (as recommended by CEOS (2021)). Further, our results suggest
that the imprecision of FIA plot locations (7.05 m standard deviation;
Table~\ref{tbl-measurement-err}) does not meaningfully impact total
uncertainty. We might infer that the fidelity of 30 m Landsat imagery is
not adequate, or that forests in NYS are not sufficiently patchy for
imprecise reference data geolocation to be meaningful in this context.
Despite its relative lack of importance in this study, reference data
uncertainty may be important in regions where only small fractions of
the tree population are represented in the data used to fit the
allometric model. Regions like the global tropics with a greater
diversity of tree species (Chave et al. 2014) or regions with a scarcity
of forest inventory plots due to remoteness or lack of resources (e.g.,
lacking a national forest inventory; McRoberts, Tomppo, and Næsset
(2010)) could yield particularly large reference data uncertainty
contributions. It is therefore prudent to weigh the precision and
representativeness (Labrière et al. 2022) of a given system of
allometric models against the practicality and feasibility of
incorporating reference data uncertainty in estimates of total variance.

In general, the framework we have developed may facilitate a range of
sensitivity analyses for different components of uncertainty and
parameters not limited to those considered here. By keeping all other
inputs constant, and increasing (or conversely removing) uncertainty
from a single component, we can begin to identify specific thresholds
beyond which more granular uncertainty contributions overwhelm the total
uncertainty estimate. Although these kinds of analyses may be simple in
theory, they are computationally expensive, as testing each component
requires a new 1000 iteration bootstrap procedure. With further
investments and developments in computational infrastructure such
analyses may become more tractable. Nevertheless, we recommend this kind
of framework for iteratively exploring sources of model-based
uncertainty.

\hypertarget{communicating-uncertainty-to-map-users}{%
\subsection{Communicating uncertainty to map
users}\label{communicating-uncertainty-to-map-users}}

Although pixel-level predictions in fine-resolution maps effectively
represent spatial patterns, finer resolution is often associated with
increased uncertainty, and management decisions are made at coarser
spatial scales. Thus we expect that users will spatially aggregate
groups of pixels into units representing forest stands, ownership
parcels, or other practically relevant units for formal estimation. The
delivery of corresponding uncertainty maps (pixel-level variance) alone
fails to facilitate statistically sound estimates of aggregate
uncertainty, as pixel-level variances cannot be naively averaged within
areas of interest in the same way that AGB predictions can (Wadoux and
Heuvelink 2023). We therefore explored alternative ways to communicate
the methods and metadata required to make rigorous uncertainty estimates
for map subregions.

In this study we found that parcel-level uncertainty (SE) varied
substantially with parcel size, forest cover, and AGB density, thus
enabling the use of a natural log-log multiple regression
(Table~\ref{tbl-se-lm}; Appendix F) for accurately estimating SE for any
subregion of the AGB maps from L. K. Johnson et al. (2023). Given the
coefficients of this regression, map users need only compute the
predictors for their area of interest and substitute them into the model
formula (Equation~\ref{eq-logmod}). All of the parcel characteristics
included as predictors in the model are intended to be accessible to map
users since each can be computed with a standard GIS and a bounding
polygon for the user's area of interest.

We do not expect that this type of regression will be useful across all
landscapes, ecological domains, or modeling frameworks. However, we do
expect this type of regression to be most effective for models that
exhibit heteroscedastic residual variance, or in regions where the
majority of estimation unit sizes are relatively small. In these
situations, both the positive correlation between AGB and residual
variance and the strong inverse relationship between estimation unit
size and spatial covariance of residuals (\(\operatorname{Var_{res}}\);
Equation~\ref{eq-varres}) can be leveraged to make accurate estimates.
In regions with constant residual variance, or where the majority of
estimation units have dimensions that exceed the range of spatial
autocorrelation of residuals, this type of regression may be less
effective at estimating SE. However, we expect SEs in these
circumstances to be generally small. Further, bootstrap variance
(\(\operatorname{Var_{boot}}\); Equation~\ref{eq-varboot}) may
contribute more substantially in regions with more sparsely sampled
forest inventories (\(\operatorname{Var_{sam}}\)) or less precise
allometric models (\(\operatorname{Var_{ref}}\)). In these contexts,
alternative predictors that help describe the degree to which a
subregion is represented in the reference data or forest inventory
sample may be useful (e.g.~area of applicability information; Meyer and
Pebesma (2021)). In light of our findings, we recommend further
investigation into regression as a means to communicate uncertainty for
map subregions in a diversity of contexts.

Improving the transfer of uncertainty information to map users enables a
more rigorous application of map products in decision-support contexts.
Specifically, the estimates of SE produced in this study provide
managers and decision-makers the means to compare the outcomes of
various policies and management practices across the landscape with the
added context of confidence intervals (Figure~\ref{fig-essex-cty}). For
example, the hypothesis that forests protected by conservation easements
hold larger carbon stocks can be formally tested (McRoberts 2011).
Although this study has focused entirely on estimating uncertainties
associated with AGB stocks, the same bootstrapping approach and the
parcel-level SE estimates developed here may facilitate the estimation
of uncertainties associated with stock-changes (Esteban et al. 2020),
which can explicitly account for GHG emissions and removals attributed
to the forest sector and can be used to track forest carbon dynamics
through time.

\hypertarget{limitations}{%
\subsection{Limitations}\label{limitations}}

Although this study demonstrates rigorous uncertainty estimation, our
approach is not without limitations and assumptions. In general, there
is a degree of subjectivity involved in choosing which sources of
uncertainty are included in an assessment. Here we have omitted
uncertainty in model predictor data (Table~\ref{tbl-uncertainty-types},
type d) from our framework (CEOS 2021). The models in L. K. Johnson et
al. (2023) are largely dependent on Landsat imagery that has been
temporally smoothed, segmented, and gap-filled by Landtrendr (Kennedy,
Yang, and Cohen 2010; Kennedy, Yang, et al. 2018). Although
incorporating uncertainty associated with the geometric or radiometric
features of the raw Landsat imagery was beyond the scope of this study,
we may be able to use the Landtrendr residuals (differences between raw
spectral value and Landtrendr fitted value) to assess variability in the
imagery in future studies. In consideration of computational efficiency,
we limited the number of bootstrap iterations to 1000. McRoberts et al.
(2023) proved a formal bootstrap stopping criterion that could be
applied to ensure precise and accurate estimates of SE but also
demonstrated that 1000 is often within the range of bootstrap iterations
that can produce reliable SE estimates. In light of data availability
and project efficiency, we relied on published statistics to build
normal distributions of measurement errors rather than sampling from
empirical distributions (Table~\ref{tbl-measurement-err}). While these
assumed normal distributions are, at best, approximations of true error
distributions, we deemed them acceptable given that reference data
uncertainty is often ignored due to overall intractability. We used
LiDAR-based predictions of AGB as reference data when estimating
residual spatial autocorrelation instead of field inventory reference
data (Section~\ref{sec-spatcormeth}). Although estimates of parcel-level
SE are sensitive to the strength of residual spatial correlation
(Figure~\ref{fig-spatcorr-sens}) and there is no guarantee that
residuals computed with LiDAR-based reference data have the same spatial
properties as those computed with field inventory data, this
substitution was a practical solution to achieve the spatial density and
coverage required for estimating residual spatial autocorrelation.
Finally, other bootstrapping protocols may be more desirable within the
circumstances of our study. In our case we have blended pairs
bootstrapping (resampling plots with replacement) and residual
bootstrapping (FIA plots; measurement errors, allometric errors, and
plot location errors) within the same iterative framework (Efron and
Tibshirani 1994). Wild bootstrapping (Esteban et al. 2020; R. Y. Liu
1988) may be preferred as it preserves both the original sampling
structure and the heteroskedasticity of model residuals.

\hypertarget{conclusion}{%
\section{Conclusion}\label{conclusion}}

The proliferation of fine-resolution forest biomass (AGB) and carbon
maps has not been matched with corresponding effort in the estimation
and communication of uncertainty for small-area spatial aggregates of
map subregions. To address this shortcoming, we developed and applied
methodology to produce estimates of standard error (SE) associated with
spatial averages of AGB predictions for a stratified random sample of
ownership parcels in New York State. This represents a novel model-based
uncertainty estimation approach that included all four types of
uncertainty (Table~\ref{tbl-uncertainty-types}), that used methods
compatible with algorithmic models (machine learning or nonparametric),
and that incorporated spatial autocorrelation of residuals. Our results
suggest that residual variance, largely resulting from spatial
autocorrelation of model residuals, will be the largest source of
uncertainty in landscapes like the midwestern and northeastern United
States where forests are divided into smaller ownership units. This
highlights the importance of accounting for residual spatial
autocorrelation in small-area uncertainty estimates, and implies that
improvements in model accuracy will yield the greatest reductions to
uncertainty in these regions. We further demonstrated that a regression
model relating parcel characteristics (area, perimeter, AGB density,
forest cover) to parcel SE can provide accurate uncertainty estimation
for any map subregion with only a fraction of the computing resources
and data required to do so from first principles. Overall, our approach
and outputs can directly support future regional-scale model-based
efforts and promote transparency in spatial forest carbon accounting.

\hypertarget{acknowledgements}{%
\section{Acknowledgements}\label{acknowledgements}}

We would like to thank the USDA Forest Service FIA program for their
data sharing and cooperation, and especially Brian F. Walters for his
consultation. We would also like to thank Sean Healey for his helpful
comments, the NYS GPO for compiling and serving LiDAR data, and the NYS
Department of Environmental Conservation (\#AM11459), Office of Climate
Change for funding support. The findings and conclusions in this
publication are those of the author(s) and should not be construed to
represent any official US Department of Agriculture or US Government
determination or policy.

\newpage{}

\renewcommand{\thefigure}{A\arabic{figure}}
\renewcommand{\thetable}{A\arabic{table}}
\setcounter{figure}{0}
\setcounter{table}{0}

\hypertarget{appendix-a-component-ratio-method}{%
\subsection*{Appendix A: Component ratio
method}\label{appendix-a-component-ratio-method}}
\addcontentsline{toc}{subsection}{Appendix A: Component ratio method}

\begin{figure}

{\centering \includegraphics{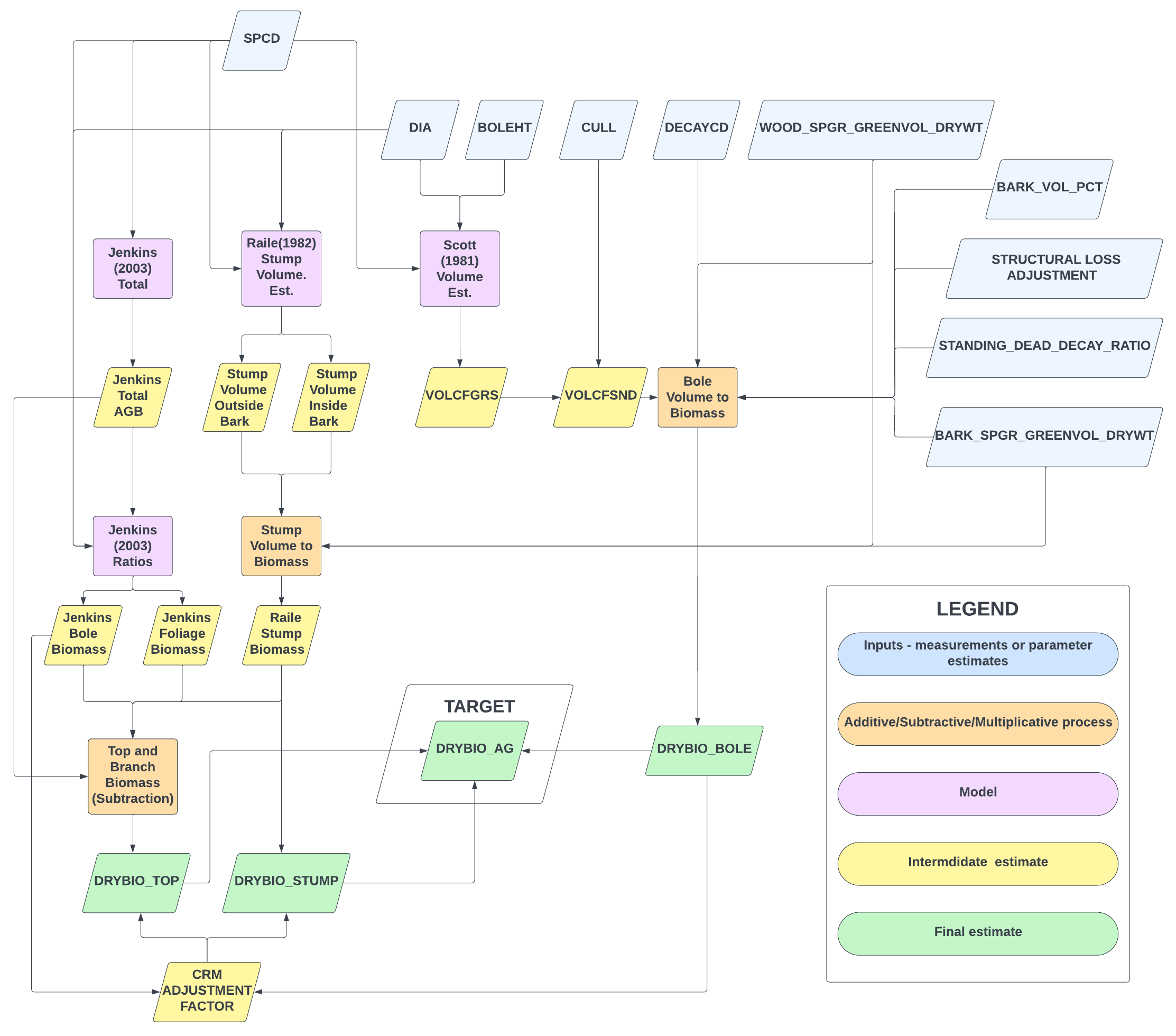}

}

\caption{Flowchart showing key elements of the component ratio method
(CRM; Woodall et al.~2011) for estimating tree-level AGB (DRYBIO\_AG)
using the FIA database. All names of inputs reflect FIA database values
as best as possible.}

\end{figure}

\newpage{}

\renewcommand{\thefigure}{B\arabic{figure}}
\renewcommand{\thetable}{B\arabic{table}}
\setcounter{figure}{0}
\setcounter{table}{0}

\hypertarget{appendix-b-legacy-tree-database}{%
\subsection*{Appendix B: Legacy tree
database}\label{appendix-b-legacy-tree-database}}
\addcontentsline{toc}{subsection}{Appendix B: Legacy tree database}

\begin{figure}

{\centering \includegraphics{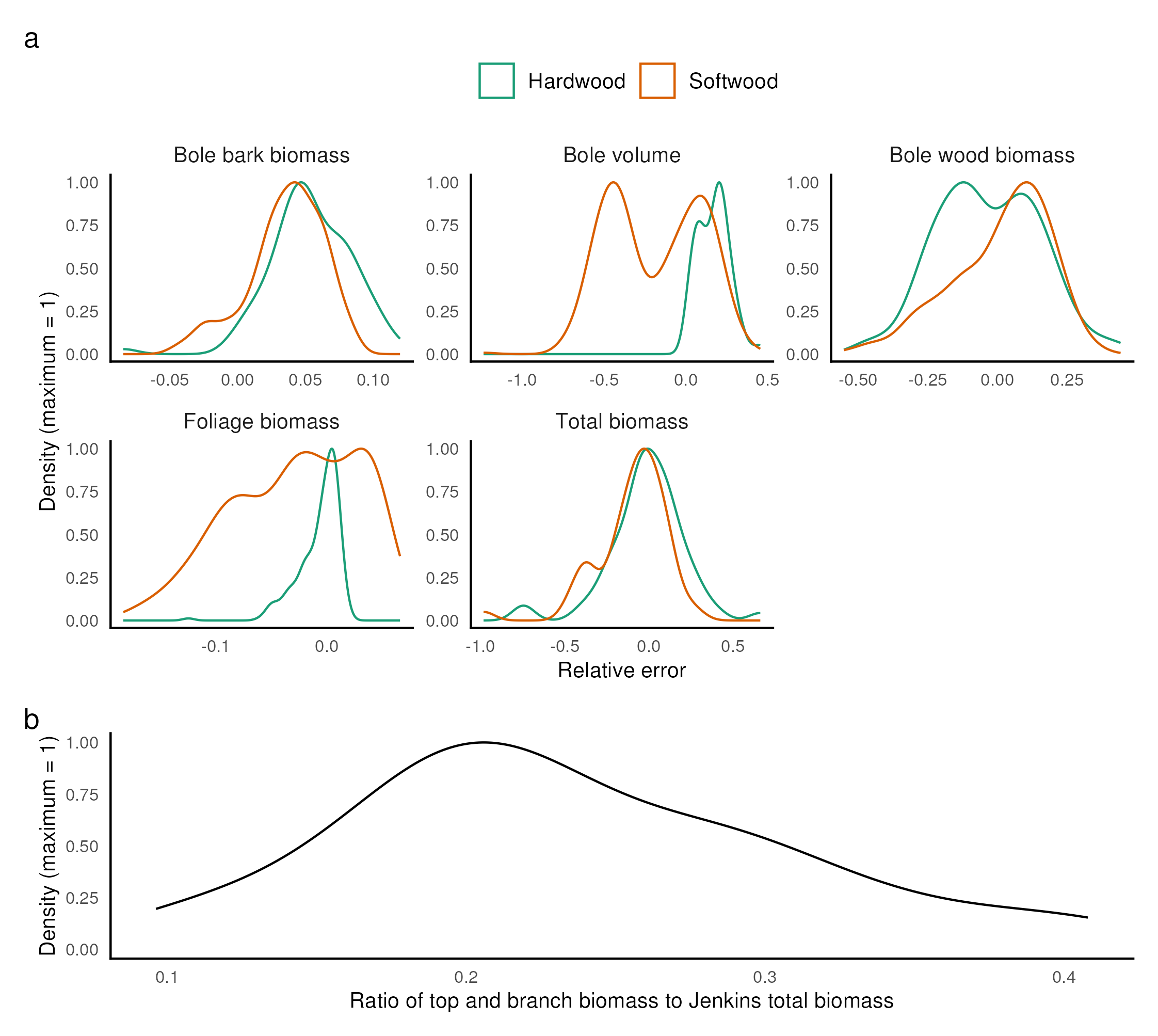}

}

\caption{\label{fig-legacydb}Smoothed frequency distributions of
allometric model uncertainty information compiled from the legacy tree
database. Distributions for all individual panels have been rescaled
separately, such that the most common occurrences within each panel are
assigned a value of 1. a) Relative residuals for five estimated tree
components. Bole volume residuals are relative to bole volume, while all
biomass residuals are relative to total tree biomass. b) Ratios for top
and branch biomass relative to total biomass.}

\end{figure}

\newpage{}

\renewcommand{\thefigure}{C\arabic{figure}}
\renewcommand{\thetable}{C\arabic{table}}
\setcounter{figure}{0}
\setcounter{table}{0}

\hypertarget{appendix-c-reference-data-uncertainty}{%
\subsection*{Appendix C: Reference data
uncertainty}\label{appendix-c-reference-data-uncertainty}}
\addcontentsline{toc}{subsection}{Appendix C: Reference data
uncertainty}

\begin{figure}

{\centering \includegraphics{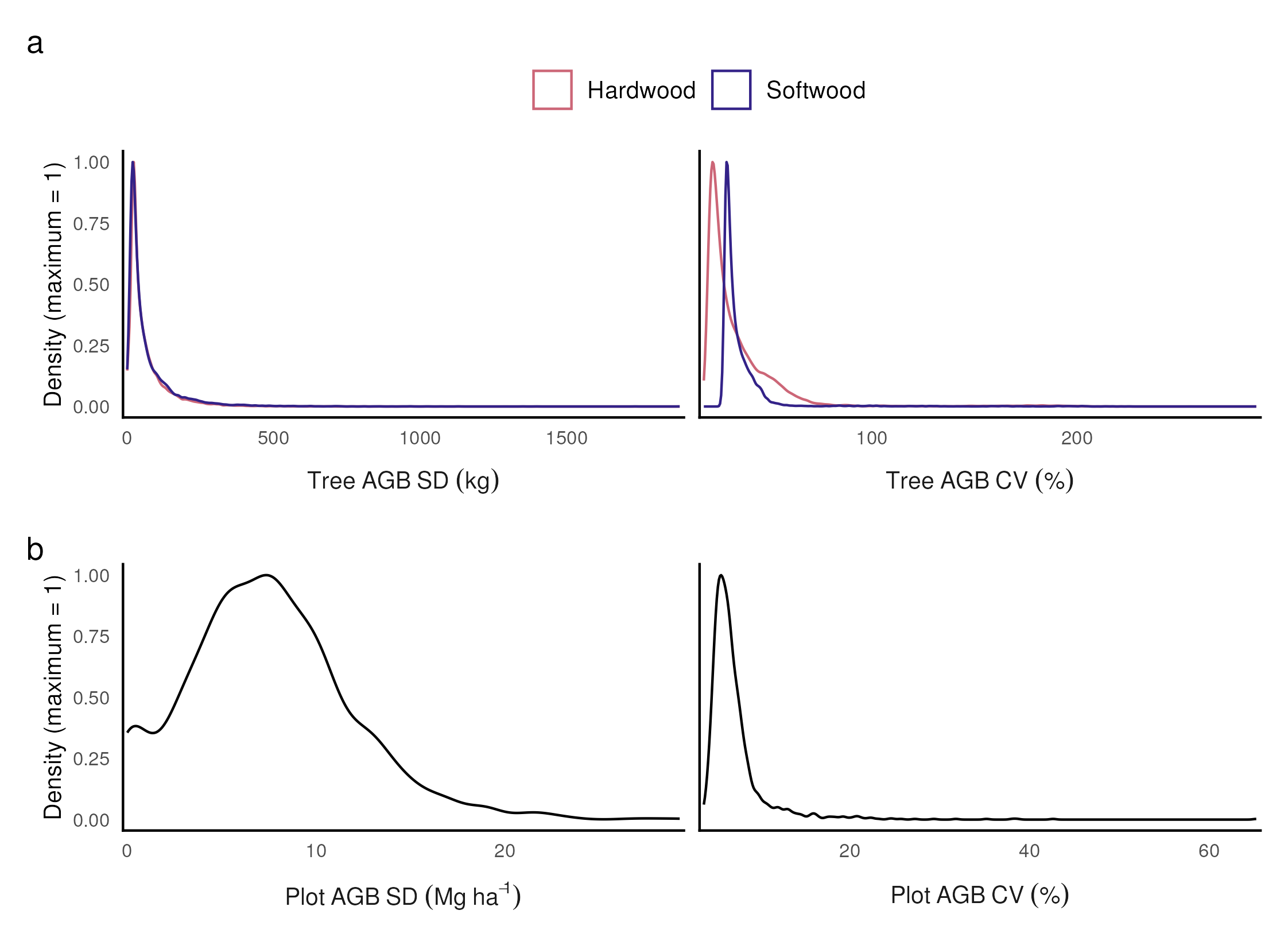}

}

\caption{\label{fig-refsd}Reference data uncertainty. Smoothed frequency
distributions estimates for individual tree (a) and plot (b) AGB
standard deviations (SD) and coefficients of variation (CV;
\(\operatorname{SD / AGB \cdot 100}\)) over the 1000 bootstrap
iterations. Distributions for individual panels have been rescaled
separately such that the most common occurrences within each panel are
assigned a value of 1.}

\end{figure}

\newpage{}

\renewcommand{\thefigure}{D\arabic{figure}}
\renewcommand{\thetable}{D\arabic{table}}
\setcounter{figure}{0}
\setcounter{table}{0}

\hypertarget{appendix-d-spatial-autocorrelation-of-residuals}{%
\subsection*{Appendix D: Spatial autocorrelation of
residuals}\label{appendix-d-spatial-autocorrelation-of-residuals}}
\addcontentsline{toc}{subsection}{Appendix D: Spatial autocorrelation of
residuals}

\begin{figure}

{\centering \includegraphics{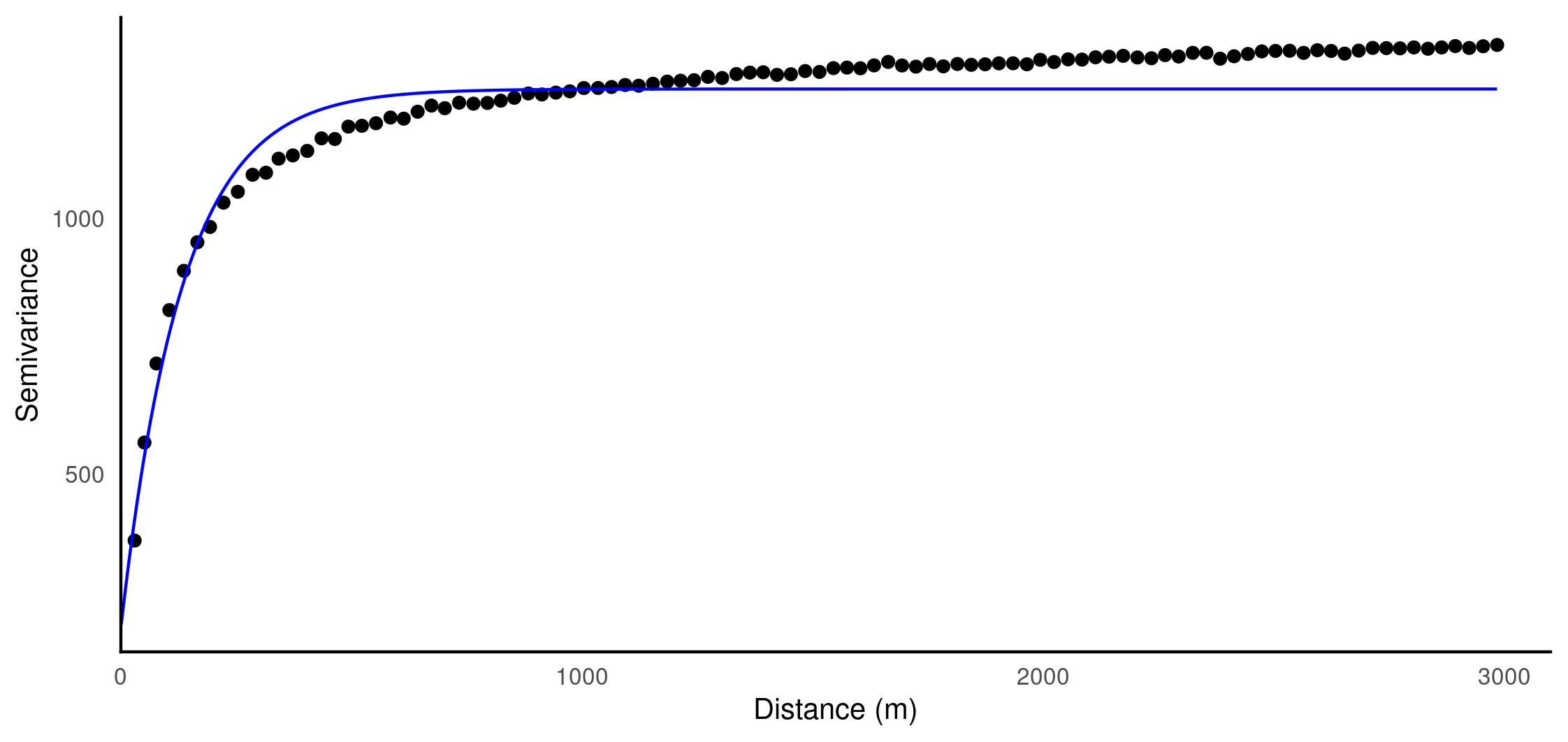}

}

\caption{\label{fig-variogram}Exponential model semivariogram (blue
line) fit to empirical semivariogram (black points) computed with a
simple random spatial sample of 500,000 residuals (linear scale). Refer
to equation 11 in the main body of the text for residual definition.}

\end{figure}

\newpage{}

\renewcommand{\thefigure}{E\arabic{figure}}
\renewcommand{\thetable}{E\arabic{table}}
\setcounter{figure}{0}
\setcounter{table}{0}

\hypertarget{appendix-e-residual-variance-model}{%
\subsection*{Appendix E: Residual variance
model}\label{appendix-e-residual-variance-model}}
\addcontentsline{toc}{subsection}{Appendix E: Residual variance model}

\begin{figure}

{\centering \includegraphics{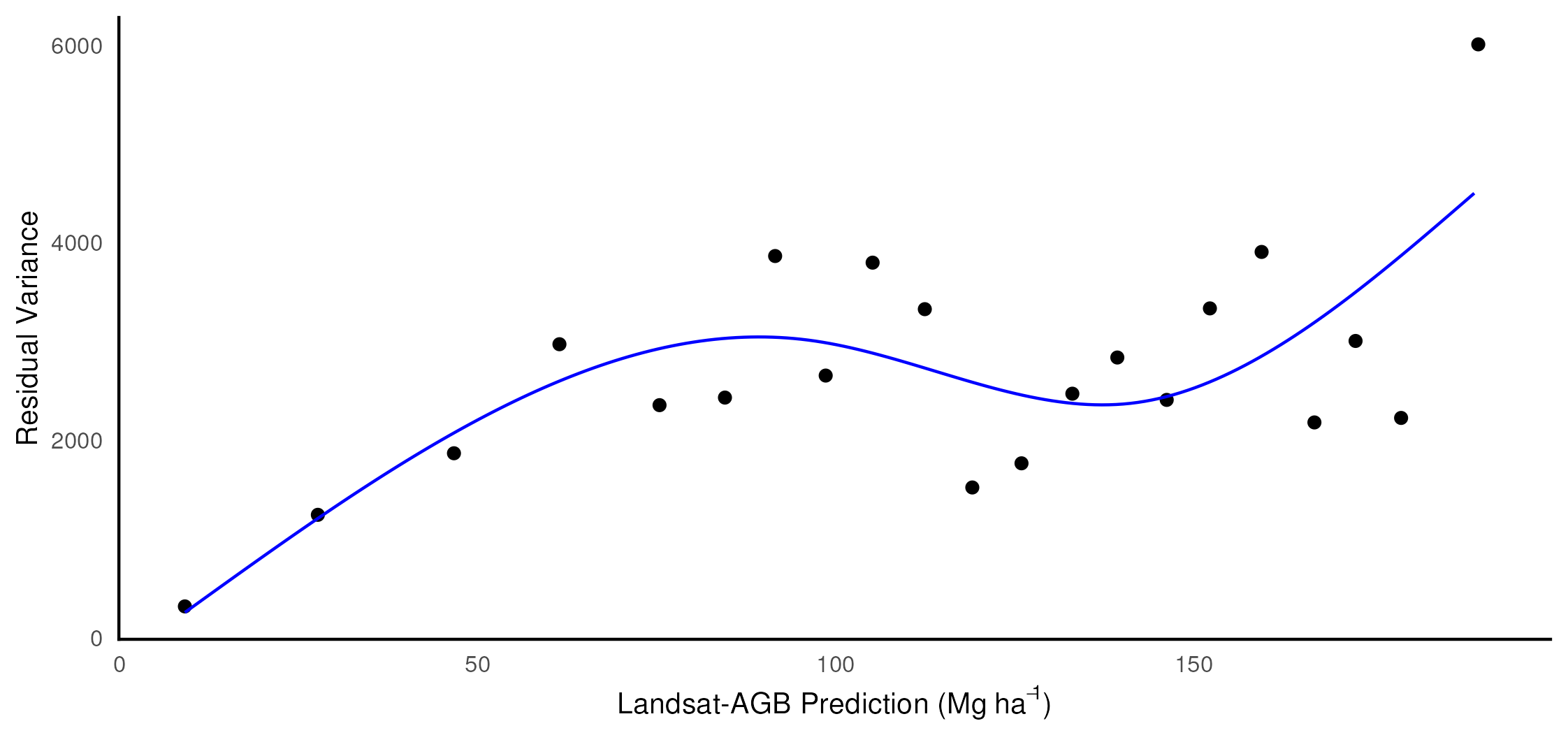}

}

\caption{\label{fig-resvarmod}Natural cubic spline residual variance
model fit (blue line) representing residual variance as a function of
AGB prediction.}

\end{figure}

\newpage{}

\renewcommand{\thefigure}{F\arabic{figure}}
\renewcommand{\thetable}{F\arabic{table}}
\setcounter{figure}{0}
\setcounter{table}{0}

\hypertarget{appendix-f-modeling-estimated-standard-error-total-variance-as-a-function-of-parcel-characteristics}{%
\subsection*{Appendix F: Modeling estimated standard error (total
variance) as a function of parcel
characteristics}\label{appendix-f-modeling-estimated-standard-error-total-variance-as-a-function-of-parcel-characteristics}}
\addcontentsline{toc}{subsection}{Appendix F: Modeling estimated
standard error (total variance) as a function of parcel characteristics}

\begin{figure}

{\centering \includegraphics{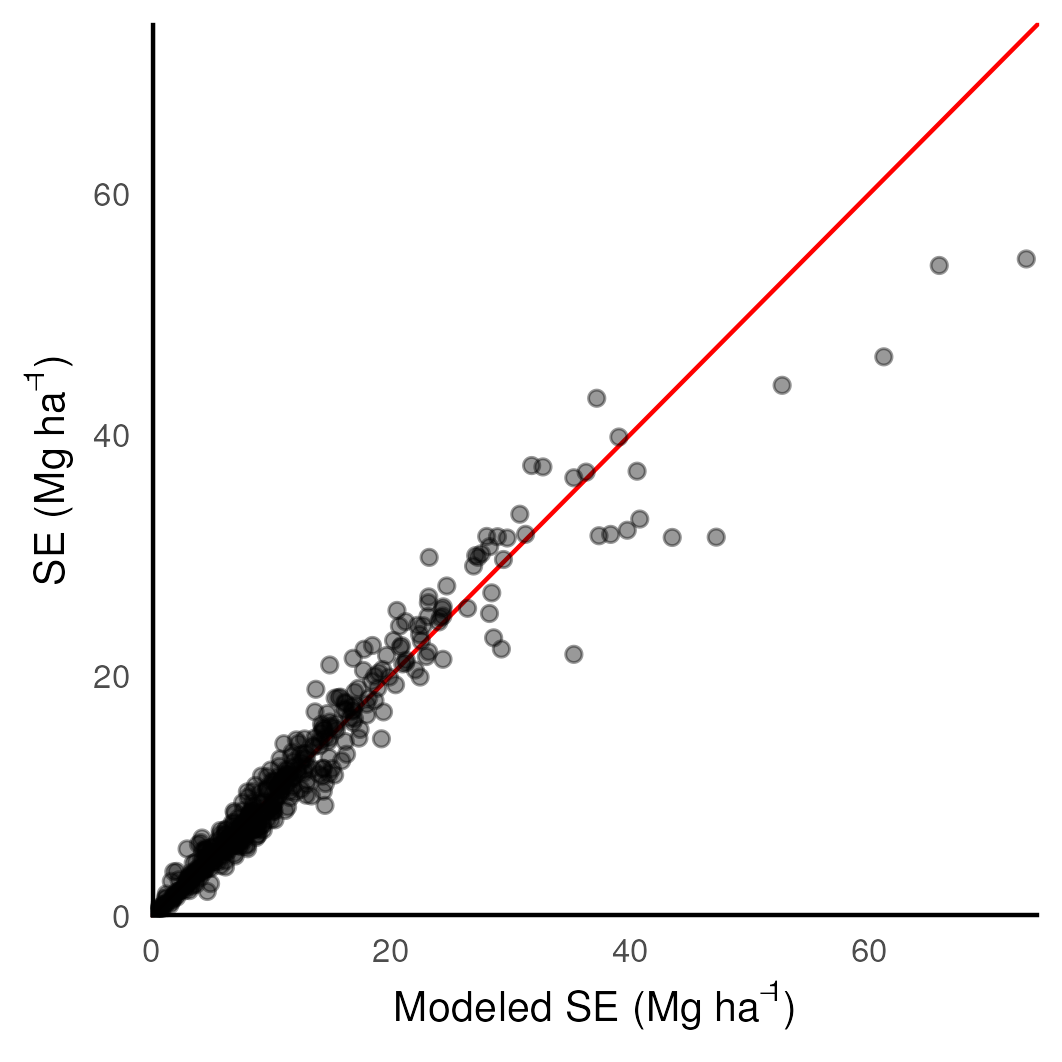}

}

\caption{\label{fig-seacc}Standard error (SE) estimated with Equations 1
and 2 vs SE estimated from a natural log-log regression with parcel
area, perimeter, AGB density, and \% forest cover as independent
variables. Each point represents a parcel from the testing partition of
the sample. 1:1 line in red.}

\end{figure}

\hypertarget{references}{%
\section*{References}\label{references}}
\addcontentsline{toc}{section}{References}

\hypertarget{refs}{}
\begin{CSLReferences}{1}{0}
\leavevmode\vadjust pre{\hypertarget{ref-Baskerville1972}{}}%
Baskerville, G. L. 1972. {``Use of Logarithmic Regression in the
Estimation of Plant Biomass.''} \emph{Canadian Journal of Forest
Research} 2 (1): 49--53. \url{https://doi.org/10.1139/x72-009}.

\leavevmode\vadjust pre{\hypertarget{ref-Baston2022}{}}%
Baston, Daniel. 2022. \emph{Exactextractr: Fast Extraction from Raster
Datasets Using Polygons}.
\url{https://CRAN.R-project.org/package=exactextractr}.

\leavevmode\vadjust pre{\hypertarget{ref-Bechtold2005}{}}%
Bechtold, William A, and Paul L Patterson. 2005. \emph{The Enhanced
Forest Inventory and Analysis Program--National Sampling Design and
Estimation Procedures}. Vol. 80. USDA Forest Service, Southern Research
Station. \url{https://doi.org/10.2737/SRS-GTR-80}.

\leavevmode\vadjust pre{\hypertarget{ref-Berger2014}{}}%
Berger, Ambros, Thomas Gschwantner, Ronald E. McRoberts, and Klemens
Schadauer. 2014. {``Effects of Measurement Errors on Individual Tree
Stem Volume Estimates for the Austrian National Forest Inventory.''}
\emph{Forest Science} 60 (1): 14--24.
\url{https://doi.org/10.5849/forsci.12-164}.

\leavevmode\vadjust pre{\hypertarget{ref-Breidenbach2014}{}}%
Breidenbach, Johannes, Clara Antón-Fernández, Hans Petersson, Ronald E.
McRoberts, and Rasmus Astrup. 2014. {``Quantifying the Model-Related
Variability of Biomass Stock and Change Estimates in the Norwegian
National Forest Inventory.''} \emph{Forest Science} 60 (1): 25--33.
\url{https://doi.org/10.5849/forsci.12-137}.

\leavevmode\vadjust pre{\hypertarget{ref-Breidenbach2016}{}}%
Breidenbach, Johannes, Ronald E. McRoberts, and Rasmus Astrup. 2016.
{``Empirical Coverage of Model-Based Variance Estimators for Remote
Sensing Assisted Estimation of Stand-Level Timber Volume.''}
\emph{Remote Sensing of Environment} 173 (February): 274--81.
\url{https://doi.org/10.1016/j.rse.2015.07.026}.

\leavevmode\vadjust pre{\hypertarget{ref-Breiman2001b}{}}%
Breiman, Leo. 2001a. {``Random Forests.''} \emph{Machine Learning} 45
(1): 5--32. \url{https://doi.org/10.1023/A:1010933404324}.

\leavevmode\vadjust pre{\hypertarget{ref-Breiman2001a}{}}%
---------. 2001b. {``Statistical Modeling: The Two Cultures (with
Comments and a Rejoinder by the Author).''} \emph{Statistical Science}
16 (3). \url{https://doi.org/10.1214/ss/1009213726}.

\leavevmode\vadjust pre{\hypertarget{ref-Brown2020}{}}%
Brown, Jesslyn F., Heather J. Tollerud, Christopher P. Barber, Qiang
Zhou, John L. Dwyer, James E. Vogelmann, Thomas R. Loveland, et al.
2020. {``{Lessons learned implementing an operational continuous United
States national land change monitoring capability: The Land Change
Monitoring, Assessment, and Projection (LCMAP) approach}.''}
\emph{Remote Sensing of Environment} 238: 111356.
\url{https://doi.org/10.1016/j.rse.2019.111356}.

\leavevmode\vadjust pre{\hypertarget{ref-IPCC2019}{}}%
Buendia, E, K Tanabe, A Kranjc, J Baasansuren, M Fukuda, S Ngarize, A
Osako, Y Pyrozhenko, P Shermanau, and S Federici. 2019. {``Refinement to
the 2006 IPCC Guidelines for National Greenhouse Gas Inventories.''}
\emph{IPCC: Geneva, Switzerland} 5: 194.

\leavevmode\vadjust pre{\hypertarget{ref-Butler2016}{}}%
Butler, Brett J., Jaketon H. Hewes, Brenton J. Dickinson, Kyle
Andrejczyk, Sarah M. Butler, and Marla Markowski-Lindsay. 2016.
{``Family Forest Ownerships of the United States, 2013: Findings from
the USDA Forest Service's National Woodland Owner Survey.''}
\emph{Journal of Forestry} 114 (6): 638--47.
\url{https://doi.org/10.5849/jof.15-099}.

\leavevmode\vadjust pre{\hypertarget{ref-CEOS}{}}%
CEOS. 2021. {``{Aboveground Woody Biomass Product Validation Good
Practices Protocol}.''}
\url{https://doi.org/10.5067/DOC/CEOSWGCV/LPV/AGB.001}.

\leavevmode\vadjust pre{\hypertarget{ref-Chave2014}{}}%
Chave, Jérôme, Maxime Réjou‐Méchain, Alberto Búrquez, Emmanuel
Chidumayo, Matthew S. Colgan, Welington B. C. Delitti, Alvaro Duque, et
al. 2014. {``Improved Allometric Models to Estimate the Aboveground
Biomass of Tropical Trees.''} \emph{Global Change Biology} 20 (10):
3177--90. \url{https://doi.org/10.1111/gcb.12629}.

\leavevmode\vadjust pre{\hypertarget{ref-Chen2015}{}}%
Chen, Qi, Gaia Vaglio Laurin, and Riccardo Valentini. 2015.
{``Uncertainty of Remotely Sensed Aboveground Biomass over an African
Tropical Forest: Propagating Errors from Trees to Plots to Pixels.''}
\emph{Remote Sensing of Environment} 160 (April): 134--43.
\url{https://doi.org/10.1016/j.rse.2015.01.009}.

\leavevmode\vadjust pre{\hypertarget{ref-Chen2016}{}}%
Chen, Qi, Ronald E. McRoberts, Changwei Wang, and Philip J. Radtke.
2016. {``Forest Aboveground Biomass Mapping and Estimation Across
Multiple Spatial Scales Using Model-Based Inference.''} \emph{Remote
Sensing of Environment} 184 (October): 350--60.
\url{https://doi.org/10.1016/j.rse.2016.07.023}.

\leavevmode\vadjust pre{\hypertarget{ref-Cooke2000}{}}%
Cooke, William H. 2000. {``Forest/Non-Forest Stratification in Georgia
with Landsat Thematic Mapper Data.''} In \emph{McRoberts, Ronald e.;
Reams, Gregory a.; van Deusen, Paul c., Eds. Proceedings of the First
Annual Forest Inventory and Analysis Symposium; Gen. Tech. Rep. NC-213.
St. Paul, MN: US Department of Agriculture, Forest Service, North
Central Research Station: 28-30}.

\leavevmode\vadjust pre{\hypertarget{ref-Cortes1995}{}}%
Cortes, Corinna, and Vladimir Vapnik. 1995. {``Support-Vector
Networks.''} \emph{Machine Learning} 20 (3): 273--97.
\url{https://doi.org/10.1007/bf00994018}.

\leavevmode\vadjust pre{\hypertarget{ref-Domke2011}{}}%
Domke, Grant M, Christopher W Woodall, and James E Smith. 2011.
{``Accounting for Density Reduction and Structural Loss in Standing Dead
Trees: Implications for Forest Biomass and Carbon Stock Estimates in the
United States.''} \emph{Carbon Balance and Management} 6 (1).
\url{https://doi.org/10.1186/1750-0680-6-14}.

\leavevmode\vadjust pre{\hypertarget{ref-Dubayah2020}{}}%
Dubayah, Ralph, James Bryan Blair, Scott Goetz, Lola Fatoyinbo, Matthew
Hansen, Sean Healey, Michelle Hofton, et al. 2020. {``The Global
Ecosystem Dynamics Investigation: High-Resolution Laser Ranging of the
Earth's Forests and Topography.''} \emph{Science of Remote Sensing} 1
(June): 100002. \url{https://doi.org/10.1016/j.srs.2020.100002}.

\leavevmode\vadjust pre{\hypertarget{ref-Dyer2006}{}}%
Dyer, James M. 2006. {``Revisiting the Deciduous Forests of Eastern
North America.''} \emph{BioScience} 56 (4): 341--52.
\url{https://doi.org/10.1641/0006-3568(2006)56\%5B341:RTDFOE\%5D2.0.CO;2}.

\leavevmode\vadjust pre{\hypertarget{ref-NAIP2019}{}}%
Earth Resources Observation And Science (EROS) Center. 2019. {``National
Agriculture Imagery Program (NAIP).''} U.S. Geological Survey.
\url{https://doi.org/10.5066/F7QN651G}.

\leavevmode\vadjust pre{\hypertarget{ref-Efron2020}{}}%
Efron, Bradley. 2020. {``Prediction, Estimation, and Attribution.''}
\emph{Journal of the American Statistical Association} 115 (530):
636--55. \url{https://doi.org/10.1080/01621459.2020.1762613}.

\leavevmode\vadjust pre{\hypertarget{ref-Efron1994}{}}%
Efron, Bradley, and Robert J Tibshirani. 1994. \emph{An Introduction to
the Bootstrap}. CRC press.

\leavevmode\vadjust pre{\hypertarget{ref-IPCC2006}{}}%
Eggleston, H S, L Buendia, K Miwa, T Ngara, and K Tanabe. 2006. {``2006
IPCC Guidelines for National Greenhouse Gas Inventories.''}

\leavevmode\vadjust pre{\hypertarget{ref-Esteban2020}{}}%
Esteban, Jessica, Ronald E. McRoberts, Alfredo Fernández-Landa, José
Luis Tomé, and Miguel Marchamalo. 2020. {``A Model-Based Volume
Estimator That Accounts for Both Land Cover Misclassification and Model
Prediction Uncertainty.''} \emph{Remote Sensing} 12 (20): 3360.
\url{https://doi.org/10.3390/rs12203360}.

\leavevmode\vadjust pre{\hypertarget{ref-Freedman1981}{}}%
Freedman, David A. 1981. {``Bootstrapping Regression Models.''}
\emph{The Annals of Statistics} 9 (6): 1218--28.
\url{https://doi.org/10.1214/aos/1176345638}.

\leavevmode\vadjust pre{\hypertarget{ref-Friedman2002}{}}%
Friedman, Jerome H. 2002. {``Stochastic Gradient Boosting.''}
\emph{Computational Statistics and Data Analysis} 38 (4): 367--78.
\url{https://doi.org/10.1016/S0167-9473(01)00065-2}.

\leavevmode\vadjust pre{\hypertarget{ref-Graler2016}{}}%
Gräler, Benedikt, Edzer Pebesma, and Gerard Heuvelink. 2016.
{``{Spatio-Temporal Interpolation using gstat}.''} \emph{{The R
Journal}} 8 (1): 204--18. \url{https://doi.org/10.32614/RJ-2016-014}.

\leavevmode\vadjust pre{\hypertarget{ref-Gray2012}{}}%
Gray, Andrew N, Thomas J Brandeis, John D Shaw, William H McWilliams,
and Patrick Miles. 2012. {``{Forest Inventory and Analysis Database of
the United States of America (FIA)}.''} \emph{Biodiversity and Ecology}
4: 225--31. \url{https://doi.org/10.7809/b-e.00079}.

\leavevmode\vadjust pre{\hypertarget{ref-Harmon2011}{}}%
Harmon, Mark E., Christopher W. Woodall, Becky Fasth, Jay Sexton, and
Misha. Yatkov. 2011. \emph{Differences Between Standing and Downed Dead
Tree Wood Density Reduction Factors: A Comparison Across Decay Classes
and Tree Species}. U.S. Department of Agriculture, Forest Service,
Northern Research Station. \url{https://doi.org/10.2737/nrs-rp-15}.

\leavevmode\vadjust pre{\hypertarget{ref-spline}{}}%
Hastie, Trevor J. 1992. {``Generalized Additive Models.''} In
\emph{Chapter 7 of Statistical Models in s}, edited by DM Bates, JM
Chambers, and T Hastie. Routledge.

\leavevmode\vadjust pre{\hypertarget{ref-terra}{}}%
Hijmans, Robert J. 2023. \emph{Terra: Spatial Data Analysis}.
\url{https://CRAN.R-project.org/package=terra}.

\leavevmode\vadjust pre{\hypertarget{ref-Hoppus2005}{}}%
Hoppus, Michael, and Andrew Lister. 2005. {``The Status of Accurately
Locating Forest Inventory and Analysis Plots Using the Global
Positioning System.''} In \emph{Proceedings of the Seventh Annual Forest
Inventory and Analysis Symposium}.
\url{https://www.fs.usda.gov/research/treesearch/17040}.

\leavevmode\vadjust pre{\hypertarget{ref-Huang2019}{}}%
Huang, Wenli, Katelyn Dolan, Anu Swatantran, Kristofer Johnson, Hao
Tang, Jarlath O'Neil-Dunne, Ralph Dubayah, and George Hurtt. 2019.
{``{High-resolution mapping of aboveground biomass for forest carbon
monitoring system in the Tri-State region of Maryland, Pennsylvania and
Delaware, {USA}}.''} \emph{Environmental Research Letters} 14 (9):
095002. \url{https://doi.org/10.1088/1748-9326/ab2917}.

\leavevmode\vadjust pre{\hypertarget{ref-Hudak2020}{}}%
Hudak, Andrew T, Patrick A Fekety, Van R Kane, Robert E Kennedy, Steven
K Filippelli, Michael J Falkowski, Wade T Tinkham, et al. 2020. {``A
Carbon Monitoring System for Mapping Regional, Annual Aboveground
Biomass Across the Northwestern {USA}.''} \emph{Environmental Research
Letters} 15 (9): 095003. \url{https://doi.org/10.1088/1748-9326/ab93f9}.

\leavevmode\vadjust pre{\hypertarget{ref-Jenkins2003}{}}%
Jenkins, Jennifer C, David C Chojnacky, Linda S Heath, and Richard A
Birdsey. 2003. {``National-Scale Biomass Estimators for United States
Tree Species.''} \emph{Forest Science} 49 (1): 12--35.
\url{https://academic.oup.com/forestscience/article/49/1/12/4617214}.

\leavevmode\vadjust pre{\hypertarget{ref-Johnson2015}{}}%
Johnson, Kristofer D., Richard Birdsey, Jason Cole, Anu Swatantran,
Jarlath O'Neil-Dunne, Ralph Dubayah, and Andrew Lister. 2015.
{``Integrating LIDAR and Forest Inventories to Fill the Trees Outside
Forests Data Gap.''} \emph{Environmental Monitoring and Assessment} 187
(10). \url{https://doi.org/10.1007/s10661-015-4839-1}.

\leavevmode\vadjust pre{\hypertarget{ref-Johnson2022}{}}%
Johnson, Lucas K., Michael J. Mahoney, Eddie Bevilacqua, Stephen V.
Stehman, Grant M. Domke, and Colin M. Beier. 2022. {``Fine-Resolution
Landscape-Scale Biomass Mapping Using a Spatiotemporal Patchwork of
{LiDAR} Coverages.''} \emph{International Journal of Applied Earth
Observation and Geoinformation} 114 (November): 103059.
\url{https://doi.org/10.1016/j.jag.2022.103059}.

\leavevmode\vadjust pre{\hypertarget{ref-Johnson2023}{}}%
Johnson, Lucas K., Michael J. Mahoney, Madeleine L. Desrochers, and
Colin M. Beier. 2023. {``Mapping Historical Forest Biomass for
Stock-Change Assessments at Parcel to Landscape Scales.''} \emph{Forest
Ecology and Management} 546 (October): 121348.
\url{https://doi.org/10.1016/j.foreco.2023.121348}.

\leavevmode\vadjust pre{\hypertarget{ref-kernlab}{}}%
Karatzoglou, Alexandros, Alex Smola, Kurt Hornik, and Achim Zeileis.
2004. {``Kernlab -- an {S4} Package for Kernel Methods in {R}.''}
\emph{Journal of Statistical Software} 11 (9): 1--20.
\url{https://doi.org/10.18637/jss.v011.i09}.

\leavevmode\vadjust pre{\hypertarget{ref-Kattenborn2021}{}}%
Kattenborn, Teja, Jens Leitloff, Felix Schiefer, and Stefan Hinz. 2021.
{``Review on Convolutional Neural Networks (CNN) in Vegetation Remote
Sensing.''} \emph{ISPRS Journal of Photogrammetry and Remote Sensing}
173 (March): 24--49.
\url{https://doi.org/10.1016/j.isprsjprs.2020.12.010}.

\leavevmode\vadjust pre{\hypertarget{ref-Guolin2017}{}}%
Ke, Guolin, Qi Meng, Thomas Finley, Taifeng Wang, Wei Chen, Weidong Ma,
Qiwei Ye, and Tie-Yan Liu. 2017. {``{LightGBM: A Highly Efficient
Gradient Boosting Decision Tree}.''} In \emph{Advances in Neural
Information Processing Systems}, edited by I. Guyon, U. V. Luxburg, S.
Bengio, H. Wallach, R. Fergus, S. Vishwanathan, and R. Garnett. Vol. 30.
Curran Associates, Inc.
\url{https://proceedings.neurips.cc/paper/2017/file/6449f44a102fde848669bdd9eb6b76fa-Paper.pdf}.

\leavevmode\vadjust pre{\hypertarget{ref-Kennedy2018b}{}}%
Kennedy, Robert E, Janet Ohmann, Matt Gregory, Heather Roberts, Zhiqiang
Yang, David M Bell, Van Kane, et al. 2018. {``An Empirical, Integrated
Forest Biomass Monitoring System.''} \emph{Environmental Research
Letters} 13 (2): 025004. \url{https://doi.org/10.1088/1748-9326/aa9d9e}.

\leavevmode\vadjust pre{\hypertarget{ref-Kennedy2010}{}}%
Kennedy, Robert E, Zhiqiang Yang, and Warren B. Cohen. 2010.
{``Detecting Trends in Forest Disturbance and Recovery Using Yearly
Landsat Time Series: 1. {LandTrendr} {\textemdash} Temporal Segmentation
Algorithms.''} \emph{Remote Sensing of Environment} 114 (12):
2897--2910. \url{https://doi.org/10.1016/j.rse.2010.07.008}.

\leavevmode\vadjust pre{\hypertarget{ref-Kennedy2018}{}}%
Kennedy, Robert E, Zhiqiang Yang, Noel Gorelick, Justin Braaten, Lucas
Cavalcante, Warren B. Cohen, and Sean Healey. 2018. {``Implementation of
the LandTrendr Algorithm on Google Earth Engine.''} \emph{Remote
Sensing} 10 (5). \url{https://doi.org/10.3390/rs10050691}.

\leavevmode\vadjust pre{\hypertarget{ref-Knapp2007}{}}%
Knapp, Alan K., John M. Briggs, Scott L. Collins, Steven R. Arecher, M.
Syndonia Bret-Harte, Brent E. Ewers, Debra P. Peters, et al. 2007.
{``Shrub Encroachment in North American Grasslands: Shifts in Growth
Form Dominance Rapidly Alters Control of Ecosystem Carbon Inputs.''}
\emph{Global Change Biology} 14 (3): 615--23.
\url{https://doi.org/10.1111/j.1365-2486.2007.01512.x}.

\leavevmode\vadjust pre{\hypertarget{ref-yardstick}{}}%
Kuhn, Max, Davis Vaughan, and Emil Hvitfeldt. 2023. \emph{Yardstick:
Tidy Characterizations of Model Performance}.
\url{https://CRAN.R-project.org/package=yardstick}.

\leavevmode\vadjust pre{\hypertarget{ref-KyotoProt}{}}%
{``Kyoto Protocol to the United Nations Framework Convention on Climate
Change.''} 1997. UN Treaty.
\url{https://treaties.un.org/doc/Publication/UNTS/Volume\%202303/v2303.pdf}.

\leavevmode\vadjust pre{\hypertarget{ref-Lroe2013}{}}%
L'Roe, Andrew W, and Shorna Broussard Allred. 2013. {``{Thriving or
Surviving? Forester Responses to Private Forestland Parcelization in New
York State}.''} \emph{Small-Scale Forestry} 12 (3): 353--76.
\url{https://doi.org/10.1007/s11842-012-9216-0}.

\leavevmode\vadjust pre{\hypertarget{ref-Labrire2022}{}}%
Labrière, Nicolas, Stuart J. Davies, Mathias I. Disney, Laura I.
Duncanson, Martin Herold, Simon L. Lewis, Oliver L. Phillips, et al.
2022. {``Toward a Forest Biomass Reference Measurement System for Remote
Sensing Applications.''} \emph{Global Change Biology} 29 (3): 827--40.
\url{https://doi.org/10.1111/gcb.16497}.

\leavevmode\vadjust pre{\hypertarget{ref-targets}{}}%
Landau, William Michael. 2021. {``The Targets r Package: A Dynamic
Make-Like Function-Oriented Pipeline Toolkit for Reproducibility and
High-Performance Computing.''} \emph{Journal of Open Source Software} 6
(57): 2959. \url{https://doi.org/10.21105/joss.02959}.

\leavevmode\vadjust pre{\hypertarget{ref-LeCun2015}{}}%
LeCun, Yann, Yoshua Bengio, and Geoffrey Hinton. 2015. {``Deep
Learning.''} \emph{Nature} 521 (7553): 436--44.
\url{https://doi.org/10.1038/nature14539}.

\leavevmode\vadjust pre{\hypertarget{ref-Liu1988}{}}%
Liu, Regina Y. 1988. {``Bootstrap Procedures Under Some Non-i.i.d.
Models.''} \emph{The Annals of Statistics} 16 (4).
\url{https://doi.org/10.1214/aos/1176351062}.

\leavevmode\vadjust pre{\hypertarget{ref-Liu2023}{}}%
Liu, Siyu, Martin Brandt, Thomas Nord-Larsen, Jerome Chave, Florian
Reiner, Nico Lang, Xiaoye Tong, et al. 2023. {``The Overlooked
Contribution of Trees Outside Forests to Tree Cover and Woody Biomass
Across Europe.''} \emph{Science Advances} 9 (37).
\url{https://doi.org/10.1126/sciadv.adh4097}.

\leavevmode\vadjust pre{\hypertarget{ref-Mahoney2022}{}}%
Mahoney, Michael J, Lucas K Johnson, Abigail Z Guinan, and Colin M
Beier. 2022. {``Classification and Mapping of Low-Statured Shrubland
Cover Types in Post-Agricultural Landscapes of the US Northeast.''}
\emph{International Journal of Remote Sensing} 43 (19--24): 7117--38.
\url{https://doi.org/10.1080/01431161.2022.2155086}.

\leavevmode\vadjust pre{\hypertarget{ref-Matasci2018}{}}%
Matasci, Giona, Txomin Hermosilla, Michael A. Wulder, Joanne C. White,
Nicholas C. Coops, Geordie W. Hobart, Douglas K. Bolton, Piotr
Tompalski, and Christopher W. Bater. 2018. {``Three Decades of Forest
Structural Dynamics over Canada{\textquotesingle}s Forested Ecosystems
Using Landsat Time-Series and Lidar Plots.''} \emph{Remote Sensing of
Environment} 216 (October): 697--714.
\url{https://doi.org/10.1016/j.rse.2018.07.024}.

\leavevmode\vadjust pre{\hypertarget{ref-McRoberts2011}{}}%
McRoberts, Ronald E. 2011. {``Satellite Image-Based Maps: Scientific
Inference or Pretty Pictures?''} \emph{Remote Sensing of Environment}
115 (2): 715--24. \url{https://doi.org/10.1016/j.rse.2010.10.013}.

\leavevmode\vadjust pre{\hypertarget{ref-McRoberts2016}{}}%
McRoberts, Ronald E., Qi Chen, Grant M. Domke, Göran Ståhl, Svetlana
Saarela, and James A. Westfall. 2016. {``Hybrid Estimators for Mean
Aboveground Carbon Per Unit Area.''} \emph{Forest Ecology and
Management} 378 (October): 44--56.
\url{https://doi.org/10.1016/j.foreco.2016.07.007}.

\leavevmode\vadjust pre{\hypertarget{ref-McRoberts2018}{}}%
McRoberts, Ronald E., Erik Næsset, Terje Gobakken, Gherardo Chirici,
Sonia Condés, Zhengyang Hou, Svetlana Saarela, Qi Chen, Göran Ståhl, and
Brian F. Walters. 2018. {``Assessing Components of the Model-Based Mean
Square Error Estimator for Remote Sensing Assisted Forest
Applications.''} \emph{Canadian Journal of Forest Research} 48 (6):
642--49. \url{https://doi.org/10.1139/cjfr-2017-0396}.

\leavevmode\vadjust pre{\hypertarget{ref-McRoberts2023}{}}%
McRoberts, Ronald E., Erik Næsset, Zhengyang Hou, Göran Ståhl, Svetlana
Saarela, Jessica Esteban, Davide Travaglini, Jahangir Mohammadi, and
Gherardo Chirici. 2023. {``How Many Bootstrap Replications Are Necessary
for Estimating Remote Sensing-Assisted, Model-Based Standard Errors?''}
\emph{Remote Sensing of Environment} 288 (April): 113455.
\url{https://doi.org/10.1016/j.rse.2023.113455}.

\leavevmode\vadjust pre{\hypertarget{ref-McRoberts2022}{}}%
McRoberts, Ronald E., Erik Næsset, Sassan Saatchi, and Shaun Quegan.
2022. {``Statistically Rigorous, Model-Based Inferences from Maps.''}
\emph{Remote Sensing of Environment} 279 (September): 113028.
\url{https://doi.org/10.1016/j.rse.2022.113028}.

\leavevmode\vadjust pre{\hypertarget{ref-McRobertsNFI2010}{}}%
McRoberts, Ronald E., Erkki O. Tomppo, and Erik Næsset. 2010.
{``Advances and Emerging Issues in National Forest Inventories.''}
\emph{Scandinavian Journal of Forest Research} 25 (4): 368--81.
\url{https://doi.org/10.1080/02827581.2010.496739}.

\leavevmode\vadjust pre{\hypertarget{ref-McRoberts2014}{}}%
McRoberts, Ronald E., and James A. Westfall. 2014. {``Effects of
Uncertainty in Model Predictions of Individual Tree Volume on Large Area
Volume Estimates.''} \emph{Forest Science} 60 (1): 34--42.
\url{https://doi.org/10.5849/forsci.12-141}.

\leavevmode\vadjust pre{\hypertarget{ref-Meyer2021}{}}%
Meyer, Hanna, and Edzer Pebesma. 2021. {``{Predicting into unknown
space? Estimating the area of applicability of spatial prediction
models}.''} \emph{Methods in Ecology and Evolution} 12 (9): 1620--33.
\url{https://doi.org/10.1111/2041-210x.13650}.

\leavevmode\vadjust pre{\hypertarget{ref-Nelson2010}{}}%
Nelson, Mark D., Greg C. Liknes, and Brett J. Butler. 2010. \emph{Map of
Forest Ownership in the Conterminous United States. {[}Scale 1:7, 500,
000{]}.} U.S. Department of Agriculture, Forest Service, Northern
Research Station. \url{https://doi.org/10.2737/nrs-rmap-2}.

\leavevmode\vadjust pre{\hypertarget{ref-Pan2011}{}}%
Pan, Yude, Richard A. Birdsey, Jingyun Fang, Richard Houghton, Pekka E.
Kauppi, Werner A. Kurz, Oliver L. Phillips, et al. 2011. {``A Large and
Persistent Carbon Sink in the World's Forests.''} \emph{Science} 333
(6045): 988--93. \url{https://doi.org/10.1126/science.1201609}.

\leavevmode\vadjust pre{\hypertarget{ref-ParisAgreement}{}}%
{``Paris Agreement to the United Nations Framework Convention on Climate
Change.''} 2015. UN Treaty. United Nations.
\url{https://treaties.un.org/doc/Publication/UNTS/Volume\%203156/Part/volume-3156-I-54113.pdf}.

\leavevmode\vadjust pre{\hypertarget{ref-Pebesma2004}{}}%
Pebesma, Edzer J. 2004. {``Multivariable Geostatistics in s: The Gstat
Package.''} \emph{Computers \& Geosciences} 30 (7): 683--91.
\url{https://doi.org/10.1016/j.cageo.2004.03.012}.

\leavevmode\vadjust pre{\hypertarget{ref-Penman2003}{}}%
Penman, J., M. Gytarski, T. Hiraishi, T. Krug, D. Kruger, R. Pipatti, L.
Buendia, et al. 2003. {``Good Practice Guidance for Land Use, Land-Use
Change and Forestry.''}
\url{http://www.ipcc-nggip.iges.or.jp/public/gpglulucf/gpglulucf.htm}.

\leavevmode\vadjust pre{\hypertarget{ref-Perry2022}{}}%
Perry, Charles H, Mark V Finco, T Barry, et al. 2022. {``Forest Atlas of
the United States.''} \emph{FS-1172} 1172.

\leavevmode\vadjust pre{\hypertarget{ref-Quegan2019}{}}%
Quegan, Shaun, Thuy Le Toan, Jerome Chave, Jorgen Dall, Jean-François
Exbrayat, Dinh Ho Tong Minh, Mark Lomas, et al. 2019. {``The European
Space Agency BIOMASS Mission: Measuring Forest Above-Ground Biomass from
Space.''} \emph{Remote Sensing of Environment} 227 (June): 44--60.
\url{https://doi.org/10.1016/j.rse.2019.03.032}.

\leavevmode\vadjust pre{\hypertarget{ref-R}{}}%
R Core Team. 2023. \emph{R: A Language and Environment for Statistical
Computing}. Vienna, Austria: R Foundation for Statistical Computing.
\url{https://www.R-project.org/}.

\leavevmode\vadjust pre{\hypertarget{ref-Radtke2015}{}}%
Radtke, PJ, DM Walker, AR Weiskittel, J Frank, JW Coulston, and JA
Westfall. 2015. {``Legacy Tree Data: A National Database of Detailed
Tree Measurements for Volume, Weight, and Physical Properties.''} In
\emph{Pushing Boundaries: New Directions in Inventory Techniques and
Applications: Forest Inventory and Analysis (FIA) Symposium},
2015:8--10. \url{https://www.fs.usda.gov/research/treesearch/50166}.

\leavevmode\vadjust pre{\hypertarget{ref-Raile1982}{}}%
Raile, Gerhard K. 1982. {``Estimating Stump Volume.''} U.S. Department
of Agriculture, Forest Service, North Central Forest Experiment Station.
\url{https://doi.org/10.2737/nc-rp-224}.

\leavevmode\vadjust pre{\hypertarget{ref-Rogers1996}{}}%
Rogers, Paul. 1996. {``Disturbance Ecology and Forest Management: A
Review of the Literature.''}
\url{https://www.fs.usda.gov/rm/pubs_int/int_gtr336.pdf}.

\leavevmode\vadjust pre{\hypertarget{ref-Woodbook}{}}%
Ross, Robert J. 2021. {``Wood Handbook: Wood as an Engineering
Material.''} \url{https://www.fs.usda.gov/research/treesearch/62200}.

\leavevmode\vadjust pre{\hypertarget{ref-Saarela2016}{}}%
Saarela, Svetlana, Sören Holm, Anton Grafström, Sebastian Schnell, Erik
Næsset, Timothy G. Gregoire, Ross F. Nelson, and Göran Ståhl. 2016.
{``Hierarchical Model-Based Inference for Forest Inventory Utilizing
Three Sources of Information.''} \emph{Annals of Forest Science} 73 (4):
895--910. \url{https://doi.org/10.1007/s13595-016-0590-1}.

\leavevmode\vadjust pre{\hypertarget{ref-Saarela2018}{}}%
Saarela, Svetlana, Sören Holm, Sean Healey, Hans-Erik Andersen, Hans
Petersson, Wilmer Prentius, Paul Patterson, Erik Næsset, Timothy
Gregoire, and Göran Ståhl. 2018. {``Generalized Hierarchical Model-Based
Estimation for Aboveground Biomass Assessment Using {GEDI} and Landsat
Data.''} \emph{Remote Sensing} 10 (11): 1832.
\url{https://doi.org/10.3390/rs10111832}.

\leavevmode\vadjust pre{\hypertarget{ref-Saarela2020}{}}%
Saarela, Svetlana, André Wästlund, Emma Holmström, Alex Appiah Mensah,
Sören Holm, Mats Nilsson, Jonas Fridman, and Göran Ståhl. 2020.
{``Mapping Aboveground Biomass and Its Prediction Uncertainty Using
{LiDAR} and Field Data, Accounting for Tree-Level Allometric and {LiDAR}
Model Errors.''} \emph{Forest Ecosystems} 7 (1).
\url{https://doi.org/10.1186/s40663-020-00245-0}.

\leavevmode\vadjust pre{\hypertarget{ref-Schnell2014}{}}%
Schnell, Sebastian, Dan Altrell, Göran Ståhl, and Christoph Kleinn.
2014. {``The Contribution of Trees Outside Forests to National Tree
Biomass and Carbon Stocks---a Comparative Study Across Three
Continents.''} \emph{Environmental Monitoring and Assessment} 187 (1).
\url{https://doi.org/10.1007/s10661-014-4197-4}.

\leavevmode\vadjust pre{\hypertarget{ref-Schoeneberger2009}{}}%
Schoeneberger, Michele M. 2009. {``Agroforestry: Working Trees for
Sequestering Carbon on Agricultural Lands.''} \emph{Agroforestry
Systems} 75: 27--37. \url{https://doi.org/10.1007/s10457-008-9123-8}.

\leavevmode\vadjust pre{\hypertarget{ref-Scott1981}{}}%
Scott, Charles Thomas. 1981. \emph{Northeastern Forest Survey Revised
Cubic-Foot Volume Equations}. Vol. 304. US Department of Agriculture,
Forest Service, Northeastern Forest Experiment~\ldots.
\url{https://www.fs.usda.gov/ne/newtown_square/publications/research_notes/pdfs/scanned/OCR/ne_rn304.pdf}.

\leavevmode\vadjust pre{\hypertarget{ref-Seymour2002}{}}%
Seymour, Robert S, Alan S White, and Philip G deMaynadier. 2002.
{``Natural Disturbance Regimes in Northeastern North
America---Evaluating Silvicultural Systems Using Natural Scales and
Frequencies.''} \emph{Forest Ecology and Management} 155 (1--3):
357--67. \url{https://doi.org/10.1016/s0378-1127(01)00572-2}.

\leavevmode\vadjust pre{\hypertarget{ref-lightgbm}{}}%
Shi, Yu, Guolin Ke, Damien Soukhavong, James Lamb, Qi Meng, Thomas
Finley, Taifeng Wang, et al. 2022. \emph{Lightgbm: Light Gradient
Boosting Machine}. \url{https://github.com/Microsoft/LightGBM}.

\leavevmode\vadjust pre{\hypertarget{ref-Sprugel1983}{}}%
Sprugel, D. G. 1983. {``Correcting for Bias in Log‐transformed
Allometric Equations.''} \emph{Ecology} 64 (1): 209--10.
\url{https://doi.org/10.2307/1937343}.

\leavevmode\vadjust pre{\hypertarget{ref-Stahl2014}{}}%
Ståhl, Göran, Juha Heikkinen, Hans Petersson, Jaakko Repola, and Sören
Holm. 2014. {``Sample-Based Estimation of Greenhouse Gas Emissions from
Forests{\textemdash}a New Approach to Account for Both Sampling and
Model Errors.''} \emph{Forest Science} 60 (1): 3--13.
\url{https://doi.org/10.5849/forsci.13-005}.

\leavevmode\vadjust pre{\hypertarget{ref-Stehman2021}{}}%
Stehman, Stephen V., Bruce W. Pengra, Josephine A. Horton, and Danika F.
Wellington. 2021. {``Validation of the u.s. Geological Survey's Land
Change Monitoring, Assessment and Projection (LCMAP) Collection 1.0
Annual Land Cover Products 1985--2017.''} \emph{Remote Sensing of
Environment} 265 (November): 112646.
\url{https://doi.org/10.1016/j.rse.2021.112646}.

\leavevmode\vadjust pre{\hypertarget{ref-UNFCCC}{}}%
{``The United Nations Framework Convention on Climate Change.''} 1992.
UN Treaty.
\url{https://treaties.un.org/doc/Publication/UNTS/Volume\%201771/v1771.pdf}.

\leavevmode\vadjust pre{\hypertarget{ref-Udawatta2011}{}}%
Udawatta, Ranjith P., and Shibu Jose. 2011. {``Carbon Sequestration
Potential of Agroforestry Practices in Temperate North America.''} In
\emph{Carbon Sequestration Potential of Agroforestry Systems}, 17--42.
Springer Netherlands. \url{https://doi.org/10.1007/978-94-007-1630-8_2}.

\leavevmode\vadjust pre{\hypertarget{ref-NYForests2020}{}}%
USDA Forest Service. 2020. {``Forests of {New York}, 2019.''} U.S.
Department of Agriculture, Forest Service, Northern Research Station.
\url{https://doi.org/10.2737/fs-ru-250}.

\leavevmode\vadjust pre{\hypertarget{ref-NYForests2019}{}}%
USFS. 2020. {``Forests of New York, 2019.''} United States Department of
Agriculture, Forest Service; U.S. Department of Agriculture, Forest
Service, Northern Research Station.
\url{https://doi.org/10.2737/fs-ru-250}.

\leavevmode\vadjust pre{\hypertarget{ref-VanAuken2000}{}}%
Van Auken, O. W. 2000. {``Shrub Invasions of North American Semiarid
Grasslands.''} \emph{Annual Review of Ecology and Systematics} 31 (1):
197--215. \url{https://doi.org/10.1146/annurev.ecolsys.31.1.197}.

\leavevmode\vadjust pre{\hypertarget{ref-Wadoux2023}{}}%
Wadoux, Alexandre M. J.-C., and Gerard B. M. Heuvelink. 2023.
{``Uncertainty of Spatial Averages and Totals of Natural Resource
Maps.''} \emph{Methods in Ecology and Evolution} 14 (5): 1320--32.
\url{https://doi.org/10.1111/2041-210x.14106}.

\leavevmode\vadjust pre{\hypertarget{ref-Webster2007}{}}%
Webster, Richard, and Margaret A Oliver. 2007. \emph{Geostatistics for
Environmental Scientists}. John Wiley \& Sons.

\leavevmode\vadjust pre{\hypertarget{ref-NSVB}{}}%
Westfall, James A., John W. Coulston, Andrew N. Gray, John D. Shaw,
Philip J. Radtke, David M. Walker, Aaron R. Weiskittel, et al. 2023.
{``A National-Scale Tree Volume, Biomass, and Carbon Modeling System for
the United States.''} U.S. Department of Agriculture, Forest Service.
\url{https://doi.org/10.2737/wo-gtr-104}.

\leavevmode\vadjust pre{\hypertarget{ref-Woodall2015}{}}%
Woodall, Christopher W., John W. Coulston, Grant M. Domke, Brian F.
Walters, David N. Wear, James E. Smith, Hans-Erik Andersen, et al. 2015.
{``The u.s. Forest Carbon Accounting Framework: Stocks and Stock Change,
1990-2016.''} U.S. Department of Agriculture, Forest Service, Northern
Research Station. \url{https://doi.org/10.2737/nrs-gtr-154}.

\leavevmode\vadjust pre{\hypertarget{ref-Woodall2011}{}}%
Woodall, Christopher W., Linda S. Heath, Grant M. Domke, and Michael C.
Nichols. 2011. {``Methods and Equations for Estimating Aboveground
Volume, Biomass, and Carbon for Trees in the u.s. Forest Inventory,
2010.''} U.S. Department of Agriculture, Forest Service, Northern
Research Station. \url{https://doi.org/10.2737/nrs-gtr-88}.

\leavevmode\vadjust pre{\hypertarget{ref-ranger}{}}%
Wright, Marvin N., and Andreas Ziegler. 2017. {``{ranger}: A Fast
Implementation of Random Forests for High Dimensional Data in {C++} and
{R}.''} \emph{Journal of Statistical Software} 77 (1): 1--17.
\url{https://doi.org/10.18637/jss.v077.i01}.

\leavevmode\vadjust pre{\hypertarget{ref-Yanai2023}{}}%
Yanai, Ruth D., Alexander R. Young, John L. Campbell, James A. Westfall,
Charles J. Barnett, Gretchen A. Dillon, Mark B. Green, and Christopher
W. Woodall. 2023. {``Measurement Uncertainty in a National Forest
Inventory: Results from the Northern Region of the {USA}.''}
\emph{Canadian Journal of Forest Research} 53 (3): 163--77.
\url{https://doi.org/10.1139/cjfr-2022-0062}.

\leavevmode\vadjust pre{\hypertarget{ref-Zhu2014}{}}%
Zhu, Zhe, and Curtis E. Woodcock. 2014. {``{Continuous change detection
and classification of land cover using all available Landsat data}.''}
\emph{Remote Sensing of Environment} 144: 152--71.
\url{https://doi.org/10.1016/j.rse.2014.01.011}.

\end{CSLReferences}

\end{document}